\documentclass[a4paper,12pt]{article}

\usepackage{amsmath}
\usepackage{amsfonts}
\usepackage{amssymb}
\usepackage[nosort]{cite}
\usepackage[dvips]{graphicx}

\numberwithin{equation}{section}

\def\beq#1\eeq{\begin{equation}#1\end{equation}}
\def\bes#1\ees{\begin{equation}\begin{split}#1
               \end{split}\end{equation}}
\def\bea#1\eea{\begin{align}#1\end{align}}
            
\newcommand{\Z}{\mathbb{Z}}
\newcommand{\J}{\mathbb{J}}
\newcommand{\T}{\mathbb{T}}
\newcommand{\GG}{\mathbb{G}}

\newcommand{\bra}[1]{\langle #1|}
\newcommand{\ket}[1]{|#1\rangle}
\newcommand{\dbra}[1]{\langle\!\langle #1|}
\newcommand{\dket}[1]{|#1 \rangle\!\rangle}

\newcommand{\A}{\mathcal{A}}
\newcommand{\I}{\mathcal{I}}
\newcommand{\E}{\mathcal{E}}
\newcommand{\V}{\mathcal{V}}
\newcommand{\G}{G_\mathrm{sc}}          
\newcommand{\Gid}{G_\mathrm{id}}        
\newcommand{\g}{{\mathfrak g}}
\newcommand{\h}{{\mathfrak h}}
\newcommand{\uu}{{\mathfrak u}}
\newcommand{\s}{{\mathfrak s}}
\newcommand{\oo}{{\mathfrak o}}
\newcommand{\ttt}{{\mathfrak t}}

\DeclareMathOperator*{\dirsum}{\oplus}

\begin{document} 
\baselineskip=6mm

\begin{titlepage}
\nopagebreak
\vskip 5mm
\begin{flushright}
hep-th/0306227\\
TU-690
\end{flushright}

\vskip 20mm
\begin{center}
{\Large \textbf{Twisted boundary states in Kazama-Suzuki models}}
\vskip 15mm
Hiroshi \textsc{Ishikawa}
\footnote{\tt ishikawa@tuhep.phys.tohoku.ac.jp} 
and Taro \textsc{Tani}
\footnote{\tt tani@tuhep.phys.tohoku.ac.jp}
\vskip 5mm
\textsl{%
Department of Physics, Tohoku University \\
Sendai 980-8578, JAPAN\\}
\end{center}
\vskip 15mm                

\begin{quote}
~~~
We construct Cardy states in the Kazama-Suzuki model $G/H\times U(1)$,
which satisfy the boundary condition twisted by the automorphisms 
of the coset theory.
We classify all the automorphisms of $G/H\times U(1)$ 
induced from those of the $G$ theory.
The automorphism group contains at least a $\Z_2$ as a subgroup
corresponding to the charge conjugation.
We show that in several models there exist
extra elements other than the charge conjugation 
and that the automorphism group can be larger than $\Z_2$.
We give the explicit form of the twisted Cardy states 
which are associated with the non-trivial automorphisms.
It is shown that the resulting states preserve the N=2 
superconformal algebra.
As an illustration of our construction, we give a detailed study for 
two hermitian symmetric space models 
$SU(4)/SU(2)\times SU(2)\times U(1)$ and $SO(8)/SO(6)\times U(1)$
both at level one.
We also study the action of the level-rank duality 
on the Cardy states and find the relation with the 
exceptional Cardy states originated from a conformal
embedding.
\end{quote}

\vfill
\end{titlepage}

\section{Introduction}
The understanding of geometry at small distance
is one of the fundamental issues in string theory.
Due to the existence of the string scale,
we are forced to modify the classical notion of geometry
in the stringy regime.
D-branes are appropriate objects for studying this regime 
since they can probe the distance smaller than the string scale.
From the worldsheet point of view,
D-branes are expressed as boundary states 
in an N=2 superconformal field theory (SCFT)
describing the target space of string.
One expects that the information about the stringy geometry can be extracted 
from the study of boundary states in N=2 SCFTs,
see for example \cite{Ooguri,Brunner,DG,Hori}.
The classification of all the boundary states 
in a given SCFT is therefore a crucial step towards the
understanding of stringy geometry.

Every consistent set of boundary states in a rational CFT
should satisfy the sewing relations~\cite{Cardy,CL,L}, which include the 
so-called Cardy condition~\cite{Cardy}.
Finding a set of Cardy states, {\it i.e.}, the states satisfying the
Cardy condition, is equivalent to finding a non-negative 
integer matrix representation (NIM-rep) of the fusion 
algebra~\cite{BPPZ,Gannon}.\footnote{
It should be noted that the Cardy condition is a necessary condition 
on boundary states. Actually, there is a NIM-rep which does not give 
rise to any consistent boundary states, such as the tadpole NIM-rep
of $\s \uu (2)_{2n-1}$ (see {\it e.g.} \cite{BPPZ}).} 
In any fusion algebra, there exists at least one NIM-rep, called 
the regular NIM-rep, since the fusion coefficients themselves
form a representation of the fusion algebra.
One can construct corresponding Cardy states in the charge
conjugation or the diagonal modular invariant, which are called the regular
Cardy states~\cite{Cardy}.
However, it is in general not clear whether there are other Cardy states 
compatible with the regular ones in the same modular invariant.
One way to obtain such states is to twist the boundary condition 
by the automorphism of the chiral algebra~\cite{BFS,GG}.
(The regular Cardy states correspond to the trivial automorphism.)
It is therefore important to classify all the automorphisms of 
a given chiral algebra in order to find the Cardy states compatible 
with the regular states.

The Kazama-Suzuki models~\cite{KS,KS2} are a wide class of N=2 SCFTs 
based on the coset construction, 
which contain the N=2 minimal models as a special case.
The algebraic properties of boundary states 
such as intersection numbers have been studied in \cite{LW,Nozaki}.
However, the analysis is limited to the states
corresponding to the trivial or the charge conjugation automorphisms.
It is then natural to raise a question 
whether there are other automorphisms in the Kazama-Suzuki models.
Once we have a non-trivial automorphism in the coset chiral algebra,
we can obtain the corresponding Cardy states
by the general procedure to construct boundary states in coset CFTs,
which are developed in \cite{Ishikawa,IshikawaTani,IshikawaYamaguchi}.
(For other works on boundary states in coset theories, 
see \cite{MMS,Gawedzki,ES,Fredenhagen,Kubota,Quella,Fredenhagen2,
Fredenhagen3}.)

In this paper, we give a systematic study of twisted Cardy states
in the Kazama-Suzuki models $G/H$
\footnote{For an earlier attempt, see \cite{Stanciu}.}.
We restrict ourselves to the simplest case~\cite{Schweigert,FSnonHSS},
{\it i.e.}, the models with a single $U(1)$ factor 
in $H$ and $\text{rank}\,\g=\text{rank}\,\h$.    
We first classify automorphisms of $G/H$ induced from 
those of the $G$ theory.
Clearly, the automorphism group of the Kazama-Suzuki models
contains at least a $\Z_2$ as a subgroup,
since the charge conjugation acts on the N=2 superconformal algebra
non-trivially.
To obtain the Cardy states other than those from the charge conjugation,
therefore, the automorphism group should be larger than $\Z_2$.
We show that in several models the automorphism group
actually contains non-trivial elements other than the charge conjugation.
Among the hermitian symmetric space (HSS) models,
two series of models have non-trivial automorphism group,
namely, $SU(2n)/SU(n)\times SU(n)\times U(1)$ and $SO(2n)/SO(2n-2)\times U(1)$.
We construct the Cardy states subject to
the boundary condition twisted by the automorphisms we have found.
Since the Kazama-Suzuki models are based on 
the N=1 super Kac-Moody algebra, it is obvious that the
resulting twisted Cardy states keep the N=1 SCA.
However, there is a priori no reason that the N=2 SCA is also preserved.
We show that all the boundary conditions we have found keep the N=2 SCA. 

Some of the Kazama-Suzuki models are related via the level-rank 
duality~\cite{KS,Gepner0,LVW,Schellekens,FS3,NS}.
For example, the Grassmannian model $SU(m+n)/SU(m)\times SU(n)\times U(1)$
at level $k$ is equivalent to the model $SU(k+n)/SU(k)\times SU(n)\times U(1)$
at level $m$. 
It is interesting to examine how this duality acts on the
boundary states.
The primary fields, and hence the regular Cardy states, 
are mapped with each other under this duality.
However, the correspondence among the twisted Cardy states is not so clear,
since the automorphism group we have found is model-dependent.
As an example, we take a Grassmannian model 
$SU(4)/SU(2)\times SU(2)\times U(1)$ at level one,
which is equivalent to $SU(3)/SU(2)\times U(1)$ at level two.
The automorphism group of the former model is $\Z_2\times \Z_2$.
Therefore, we obtain non-trivial Cardy states other than 
those from the charge conjugation in the former model.
We find that the level-rank duality maps these states 
to the states in the latter model not originated from
the automorphism of the current algebra of $G$.
We show that the resulting states are described by a NIM-rep 
obtained from the conformal embedding 
$\s\uu(3)_2 \oplus \s\uu(2)_3 \subset \s\uu(6)_1$ by the procedure
developed in \cite{IshikawaTani}.

The organization of this paper is as follows.
We review in section \ref{sec:states}
the construction of boundary states in coset 
theories.
In section \ref{sec:conditions}, we determine the 
automorphism group of the Kazama-Suzuki models and 
show that the corresponding twisted boundary conditions keep the N=2 SCA.
In section \ref{sec:examples},
we present the explicit construction of twisted Cardy states
in two examples: $SU(4)/SU(2)\times SU(2)\times U(1)$ at level one 
(section \ref{sec:example1})
and $SO(8)/SO(6)\times U(1)$ at level one (section \ref{sec:example2}).
These models belong to the N=2 minimal series, for which 
all the N=2 Cardy states are known.
We check that the twisted Cardy states we have found 
together with the regular ones reproduce 
all the Cardy states for the minimal models.
The action of the level-rank duality is discussed 
in section \ref{sec:exceptional}.
Section \ref{sec:summary} is devoted to summary and discussion.
Some technical details for constructing coset boundary states are
summarized in appendix \ref{sec:twistid}.
In appendix \ref{sec:KSsuper}, we present the superfield
description of the N=2 SCA.
In appendix \ref{sec:uniqueness}, we argue the  
complex structure of the Kazama-Suzuki models.
Field identifications and selection rules~\cite{MS,Gepner,LVW,Schellekens} 
used in section \ref{sec:examples} are reviewed in appendix \ref{sec:selectid}.
The explicit form of the twisted Cardy states considered in section 
\ref{sec:example} is given in appendix \ref{sec:explicit}.


\section{Boundary states in coset theories}
\label{sec:states}

In this section, we review the construction of 
the Cardy states in coset theories developed 
in \cite{Ishikawa,IshikawaTani,IshikawaYamaguchi}.

\subsection{WZW models}
\label{sec:WZWmutual}

We begin with the WZW models since the coset theories are based on them.
The chiral algebra $\A$ of the $G$-WZW model is the affine Lie algebra $\g$.
We denote by $\I$ the set $P^k_+(\g)$ of all the 
integrable highest-weight representations of $\g$ at level $k$.
In this section, we mainly consider the bulk theory with 
the charge conjugation modular invariant
\beq
   Z= \sum_{\lambda \in \I}|\chi_{\lambda}|^2,
\label{eq:ccmodular}
\eeq
where $\chi_{\lambda}$ is the character of the representation $\lambda \in \I$.
The space ${\cal H}$ of the states in this theory is decomposed as
\beq
   {\cal H} = \dirsum_{\lambda \in \I} {\cal H}_{\lambda} \otimes 
                                       \widetilde{{\cal H}}_{\bar{\lambda}},
\label{eq:ccspec}
\eeq
where ${\cal H}_{\lambda}$ ($\widetilde{{\cal H}}_{\bar{\lambda}}$) is  
the representation space in the (anti-) chiral sector.
The case of the diagonal modular invariant will be 
considered later in this subsection.

With the presence of boundaries, the chiral and the anti-chiral sectors 
are related with each other.
This is expressed in terms of boundary conditions.
The simplest one is  
\beq
   J_n + \tilde{J}_{-n}=0,
\label{eq:ccbc}
\eeq   
where $J$ ($\tilde{J}$) is the current of the (anti-) chiral sector.
Associated with this boundary condition, we can construct the regular
Cardy states~\cite{Cardy},
\beq
   \ket{\alpha} = \sum_{\lambda \in \I} \frac{S_{\alpha \lambda}}
                                             {\sqrt{S_{0 \lambda}}}
                                             \dket{\lambda},
        \quad ~~ \alpha \in \I. 
\label{eq:regular}
\eeq 
Here $0$ is the vacuum representation and 
$\{\dket{\lambda}|\,\lambda \in \I\}$ are the 
Ishibashi states~\cite{Ishibashi} with the normalization
\beq
   \dbra{\lambda} \tilde{q}^{H_{\text{c}}/2}\dket{\mu}
                  = \delta_{\lambda \mu}\, \chi_{\lambda}(-1/\tau).
\eeq
$H_{\text{c}}=L_0+\tilde{L}_0-c/12$ and $\tilde{q}=e^{-2\pi i/\tau}$.
One can see that the spectrum of the Ishibashi states for 
the regular Cardy states is compatible with
the charge conjugation modular invariant \eqref{eq:ccmodular}.

There are other boundary states in the charge conjugation
modular invariant, called the twisted Cardy states,
which are associated with the twisted boundary condition 
\beq
   \omega (J_n) + \tilde{J}_{-n} =0.
\label{eq:twistbc}
\eeq
Here $\omega$ is an outer automorphism of the affine Lie algebra $\g$.
The labels of the Ishibashi states satisfying \eqref{eq:twistbc}
are limited to those fixed by $\omega$,
which we denote by $\I^{\omega}$,
\beq
  \I^{\omega} = \{\lambda \in \I|\, \omega (\lambda) = \lambda\}.
\label{eq:twistedIshibashi}
\eeq
Since $\omega$ acts on $\I$ non-trivially, $\I^{\omega}$ is a 
proper subset of $\I$.
Corresponding Cardy states 
$\ket{\tilde{\alpha}}_{\omega}$ are written as~\cite{BFS,GG}
\beq
  \ket{\tilde{\alpha}}_{\omega} = \sum_{\lambda \in \I^{\omega}}
                         \frac{\tilde{S}_{\tilde{\alpha}\lambda}}
                              {\sqrt{S_{0 \lambda}}}
                         \dket{\lambda}_{\omega}, 
                      \quad ~~ \tilde{\alpha} \in \tilde{\I}.
\label{eq:twistedCardy}
\eeq
$\tilde{\I}$ is the set of all the possible representations of
the twisted chiral algebra $\A^{\omega}$, an algebra generated by
the currents with the twisted boundary condition 
$J(e^{2\pi i}z) = \omega (J(z))$.
$\tilde{S}_{\tilde{\lambda}\lambda}$ is the following (transposition of) 
modular transformation matrix,
\beq
   \chi_{\lambda}^{\omega}(-1/\tau) = \sum_{\tilde{\lambda}\in \tilde{\I}}
                                     \tilde{S}_{\tilde{\lambda}\lambda}
                                     \chi_{\tilde{\lambda}}(\tau/r),
       \quad ~~ \tilde{S}\tilde{S}^{\dagger} = 1.   
\eeq
Here $\chi_{\tilde{\lambda}}$ is the character of the representation 
$\tilde{\lambda} \in \tilde{\I}$,
$r$ is the order of the automorphism $\omega$
and $\chi_{\lambda}^{\omega}$ is the twining character~\cite{FSS}.
For $\A=\g^{(1)}$, one can take as $\omega$ the diagram automorphism
of the horizontal subalgebra.
Then the twisted chiral algebra is the twisted affine Lie algebra 
$\A^{\omega}=\g^{(r)}$ and 
the twining character $\chi^{\omega}_{\lambda}$
is given by the character of the orbit Lie algebra 
$\tilde{\g}^{(r)}$~\cite{FSS} (see Table \ref{table:twistedaffine}),
\begin{table}
\begin{center}
\begin{tabular}{c|ccccc}
${\mathfrak g}^{(1)}$ 
         & $A^{(1)}_{2l}$ 
                 & $A^{(1)}_{2l-1}$
                        & $D^{(1)}_{l+1}$
                               & $E^{(1)}_6$
                                      & $D^{(1)}_4$         \\  \hline \hline 
${\mathfrak g}^{(r)}$
         & $A_{2l}^{(2)}$
                 &  $A_{2l-1}^{(2)}$
                        & $D_{l+1}^{(2)}$
                               & $E_6^{(2)}$
                                      & $D_4^{(3)}$   \\ \hline  
$\tilde{{\mathfrak g}}^{(r)}$
         & $A_{2l}^{(2)}$
                 & $D_{l+1}^{(2)}$
                        & $A_{2l-1}^{(2)}$
                               & $E_6^{(2)}$
                                      & $D_4^{(3)}$    
\end{tabular}
\end{center}
\caption{The twisted affine Lie algebras and the corresponding 
orbit Lie algebras.}
\label{table:twistedaffine}
\end{table}
\beq
    \tilde{\lambda} \in \tilde{\I} = P^k_+ (\g^{(r)}),  \quad
    \lambda \in \I^{\omega} \cong P^k_+(\tilde{\g}^{(r)}).                    
\label{eq:outerCardy}
\eeq
The twisted Cardy states are compatible with the regular 
ones \eqref{eq:regular}.
In particular, one can show
\beq
  \,_{\omega}\bra{\tilde{\alpha}} \tilde{q}^{H_{\text{c}}/2}\ket{0}_{\omega=1}
                 = \chi_{\tilde{\alpha}}(\tau/r).
\label{eq:mutual}
\eeq
This is the reason that the twisted Cardy states are labeled 
by the representations $\tilde{\I}$ of the twisted chiral algebra.

When the affine Lie algebra is not simple, 
one can consider automorphisms mixing several factors.
Suppose that the chiral algebra is a direct sum 
of some affine Lie algebra $\A$,
\beq
  \A^r = \overbrace{\A \oplus \A\oplus \cdots \oplus \A}^r.
\eeq
Clearly, the cyclic permutation $\pi$
of the factors $\A$ is an automorphism of $\A^r$.
One can use this automorphism $\pi$ to twist boundary conditions and
construct the corresponding twisted Cardy states~\cite{Recknagel}.
In this case, both of the set $\I^{\pi}$ labeling the Ishibashi states 
and the set $\tilde{\I}$ labeling the twisted Cardy states 
are identified with $\I$,
\beq
\tilde{\lambda} \in \tilde{\I} = \I,
     \quad \Lambda \in \I^{\pi} = \{(\lambda,\lambda,\ldots,\lambda)|\,
                                       \lambda \in \I\}.
\eeq
The twining character is given by 
$\chi^{\pi}_{\Lambda}(-1/\tau)= \chi_{\lambda}(-r/\tau)$ and
the modular transformation matrix $\tilde{S}$ is 
equal to that for $\A$.

Before concluding this subsection,
we comment on the construction of the Cardy states 
in the diagonal modular invariant.
The natural boundary condition for the diagonal 
modular invariant is that twisted by the charge conjugation 
$\omega_c$,  
\beq
   \omega_c(J_n) + \tilde{J}_{-n}=0.
\label{eq:dbc}
\eeq  
One can construct the Cardy states corresponding to this boundary condition
by replacing the Ishibashi states $\dket{\lambda}$ in the 
regular Cardy states \eqref{eq:regular} with $\dket{\lambda}_{\omega_c}$
satisfying the condition \eqref{eq:dbc},
\beq
   \ket{\alpha} = \sum_{\lambda \in \I} \frac{S_{\alpha \lambda}}
                                             {\sqrt{S_{0 \lambda}}}
                                             \dket{\lambda}_{\omega_c},
        \quad ~~ \alpha \in \I. 
\label{eq:dregular}
\eeq 
We also call these states the regular Cardy states in the 
case of the diagonal modular invariant.
Accordingly, in this case, we refer to the following boundary
condition as the twisted one,
\beq
   \omega \omega_c(J_n) + \tilde{J}_{-n}=0,
\eeq
where $\omega$ is an automorphism of the chiral algebra.
The corresponding twisted Cardy states take the same
form as those for the charge conjugation 
modular invariant (see eq.~\eqref{eq:twistedCardy})
except that we have to use the twisted Ishibashi states 
$\dket{\lambda}_{\omega \omega_c}$ instead of $\dket{\lambda}_{\omega}$,
\beq
  \ket{\tilde{\alpha}}_{\omega} = \sum_{\lambda \in \I^{\omega}}
                         \frac{\tilde{S}_{\tilde{\alpha}\lambda}}
                              {\sqrt{S_{0 \lambda}}}
                         \dket{\lambda}_{\omega\omega_c}, 
                      \quad ~~ \tilde{\alpha} \in \tilde{\I}.
\label{eq:dtwistedCardy}
\eeq
To summarize, the Cardy states in the diagonal modular invariant
have exactly the same form as those in the charge conjugation
modular invariant.
The difference of these two sets is only in the expression of 
the Ishibashi states,
and we can translate one set into the other by taking 
appropriate Ishibashi states.

\subsection{$G/H$ theories}
\label{sec:cosetmutual}

A coset model $G/H$ is based on an embedding of the affine
Lie algebra $\h$ into $\g$~\cite{GKO}.
A representation $\lambda$ of $\g$ is decomposed into
representations $\mu$ of $\h$ as follows,
\beq
    {\cal H}_{\lambda}^G = \dirsum_{\mu}{\cal H}_{(\lambda;\mu)}
                                        {\cal H}_{\mu}^H.
\eeq  
Hence a representation of the $G/H$ theory is labeled by 
a pair of representations of the $G$ and the $H$ 
theories.
However, not all the pairs appear in this decomposition
(selection rule) and 
more than one pairs may give the same representation
(field identification)~\cite{MS,Gepner,LVW}.
Consequently, the spectrum of the $G/H$ theory reads 
\beq
  \hat{{\cal I}} = \{ (\mu;\nu)\,|\, \mu \in \I^G,\nu\in \I^H,
                                     b_{\mu}^G(J) = b_{\nu}^H(J'),
                                   ~(J\mu;J'\nu)=(\mu;\nu),
                                   ~\forall (J/J') \in \Gid   \}.
\label{eq:cosetprimary}
\eeq  
Here $\I^G$ ($\I^H$) is the set of all the representations in the
$G$ ($H$) theory. 
The condition $b_{\mu}^G(J)=b_{\nu}^H(J')$
expresses the selection rule,
while the relation $(J\mu; J'\nu)=(\mu;\nu)$ stands for 
the field identification.
$J \mu$ is the fusion of $\mu$ with $J \in \G^G$, 
the simple current of the $G$ theory~\cite{SY,Intriligator}. 
$b_{\mu}^G(J)$ is defined as 
\beq
   b_{\mu}^G(J) = \frac{S^G_{J \mu}}{S^G_{0\mu}}  \quad \quad (J \equiv J0),
\eeq
and takes values in roots of unity.\footnote{
$b_{\mu}^G(J)$ is nothing but the exponential of the
monodromy charge, $b_{\mu}^G(J)=e^{2\pi i Q_{J}(\mu)}$~\cite{SY}.}
All the field identifications $(J/J')$ form a subgroup
of $\G^G \times \G^H$, which is called the 
identification current group $\Gid$.  
The modular transformation matrix
$\hat{S}$ of the $G/H$ theory is given by those of the $G$ and $H$ theories,
\beq
   \hat{S}_{(\mu;\nu)(\mu';\nu')} = |\Gid|\,{S^G_{\mu \mu'}}
                                                 \overline{S^H_{\nu \nu'}},
\label{eq:cosetSmatrix}
\eeq
if the identification current group $\Gid$ has no fixed points.

One can construct the regular Cardy states of the $G/H$ theory 
in the same way as the WZW models,
\beq
   \ket{(\alpha;\beta)} = \sum_{(\lambda;\mu) \in \hat{\I}}
                          \frac{\hat{S}_{(\alpha;\beta)(\lambda;\mu)}}
                               {\sqrt{\hat{S}_{(0;0)(\lambda;\mu)}}}
                               \dket{(\lambda;\mu)},  \quad ~~
      (\alpha;\beta) \in \hat{\I}.  
\eeq
As we have seen in the previous subsection,
if there are automorphisms of the chiral algebra,
one can twist boundary conditions and construct 
the corresponding twisted Cardy states which coexist with 
the regular ones.
Therefore, it is important to know what kind of automorphisms 
exist in the $G/H$ theory. 
As is discussed in \cite{Ishikawa,IshikawaYamaguchi},
an automorphism $\omega^{\g}\in \text{Aut}(\g)$ of the $G$ theory 
induces an automorphism $\hat{\omega}$ of the $G/H$ theory
if $\omega^{\g}$ can be restricted on $\h$, 
{\it i.e.}, $\omega^{\g}(\h)=\h$.
(In general, not all automorphisms of the coset theory are
of this form. See section \ref{sec:exceptional}.)
Here we denote by $\text{Aut}(\g)$ the group of automorphisms of $\g$.
$\text{Aut}(\g)$ takes the following form
\beq
  \text{Aut}(\g) = \text{Ad}(G) \rtimes D(\g),
\eeq
where $\text{Ad}(G)$ is the group of inner automorphisms
$X \mapsto \text{Ad}(g) X = g X g^{-1} \, (g \in G)$ and $D(\g)$ is the group
of diagram automorphisms.
Clearly, the automorphisms that keep $\h \subset \g$ invariant
form a subgroup of $\text{Aut}(\g)$,
which we denote by $\text{Aut}_{\h}(\g)$
\beq
   \text{Aut}_{\h}(\g) = \{\omega^{\g}\in \text{Aut}(\g)\,|\,
                                 \omega^{\g}(\h) =\h\}.
\eeq
Each element $\omega^{\g} \in \text{Aut}_{\h}(\g)$ induces
an automorphism $\hat{\omega}$ of the $G/H$ theory defined as follows,
\beq
   \hat{\omega}:~ (\mu;\nu) \mapsto 
                  (\omega^{\g}(\mu);\omega^{\h}(\nu)),\quad 
                                                (\mu;\nu) \in \hat{\I},
\label{eq:autobranch}
\eeq
where $\omega^{\h}=\omega^{\g}|_{\h}$.

Not all the automorphisms $\omega^{\g} \in \text{Aut}_{\h}(\g)$
induce non-trivial automorphisms in the coset theory.
Let us take an element $h \in H$ and consider its adjoint action
on the currents, $J \mapsto \text{Ad}(h) J  = h J h^{-1}$.
Although  $\omega^{\g}=\text{Ad}(h)$ is a non-trivial element of 
$\text{Aut}_{\h}(\g)$,
the corresponding automorphism $\hat{\omega}$ 
acts trivially on the coset theory.
In order to see this, let us note that the character $\chi^G_{\mu}$ of 
$\mu \in \I^G$ 
with the insertion of $h \in H$ has the following branching rule
\beq
  \chi^G_{\mu}(\tau,z) = \sum_{\nu \in {\mathcal I}^H} 
                       \chi_{(\mu;\nu)}(\tau) 
                       \chi^H_{\nu}(\tau,z)  ,
\eeq
where $z$ is a weight of $\h$ depending 
only on the conjugacy class of $h \in H$.
Since $z$ is a weight of $\h$, the branching function $\chi_{(\mu;\nu)}$ 
does not depend on $z$.
The insertion of $h$ corresponds to the twist by $\omega^{\g}=\text{Ad}(h)$
and the independence of $\chi_{(\mu;\nu)}$ on $z$ shows that
$\text{Ad}(h)$ acts trivially on the coset theory. 

All the conjugations $\text{Ad}(h)$ by the elements $h$ of $H$
form a normal subgroup $\text{Ad}(H) \vartriangleleft \text{Aut}_\h(\g)$.
Since $\text{Ad}(H)$ acts on the $G/H$ theory trivially,
we define the group of induced automorphisms of the $G/H$ theory
as the quotient of $\text{Aut}_\h(\g)$ by $\text{Ad}(H)$
\beq
  \text{Aut}(G/H) = \text{Aut}_\h(\g) / \text{Ad}(H) . 
\label{eq:AutGH}
\eeq
Let $[\omega^{\g}]$ ($\omega^{\g} \in \text{Aut}_\h(\g)$) 
be an element of $\text{Aut}(G/H)$.
We can take the representative $\omega^{\g}$ such that 
$\omega^{\g}$ maps the  Cartan subalgebra $\ttt_{\h}$ of $\h$ to itself.
In general, $\omega^{\g} \in \text{Aut}_\h(\g)$ maps $\ttt_{\h}$
to another maximal Abelian subalgebra of $\h$.
However, all the maximal Abelian subalgebras of $\h$
are conjugate to each other by the action of $\text{Ad}(H)$.
Therefore, we can choose an element $h \in H$ so that the
following equation holds,
\beq
   \omega^{\g}(\ttt_\h) = \text{Ad}(h) (\ttt_\h),
\eeq  
which means that $\text{Ad}(h^{-1}) \omega^{\g}$ maps
$\ttt_{\h}$ to itself.
Since we consider the quotient by $\text{Ad}(H)$,
$\omega^{\g}$ and $\text{Ad}(h^{-1}) \omega^{\g}$ determine 
the same element in $\text{Aut}(G/H)$,
$
  [\omega^{\g}]=[\text{Ad}(h^{-1}) \omega^{\g}].
$
To determine the group $\text{Aut}(G/H)$, 
it is hence sufficient to consider the automorphisms that
map $\ttt_{\h}$ to itself.

Furthermore, when $\text{rank}\, \g = \text{rank} \, \h$,
we can express the representatives in more concise form.
In this case, the Cartan subalgebras of $\g$ and $\h$ coincide 
$\ttt_\h=\ttt_\g$, and we can take the representative of 
$[\omega^{\g}] \in \text{Aut}(G/H)$ from automorphisms 
that map $\ttt_{\g}$ to itself, which form the group 
\beq
\text{Aut}_{\ttt_\g}(\g)=\{\omega^{\g} \in \text{Aut}(\g)\,|\, 
                      \omega^{\g}(\ttt_\g) =\ttt_\g \}.
\eeq
Therefore, the group of induced automorphism of the G/H theory 
(\ref{eq:AutGH}) can be written as
\beq
 \text{Aut}(G/H) =  \text{Aut}_{\ttt_\g}(\g)\cap \text{Aut}_{\h}(\g)
 / \text{Aut}_{\ttt_\g}(\g)\cap \text{Ad}(H).
\label{eq:AutGH1}
\eeq
It is known~\cite{Helgason} that there is a homomorphism $\varphi$ from 
$\text{Aut}_{\ttt_\g}(\g)$ 
{\it onto} the automorphism group $\text{Aut}(Q^{\g})$
of the root lattice $Q^\g$ of $\g$
\beq
    \varphi :~ \text{Aut}_{\ttt_\g}(\g) \rightarrow \text{Aut}(Q^{\g}).
\eeq
The kernel of $\varphi$ is $\text{Ad}(T^{G})$,
where $T^G$ is the Cartan subgroup of $G$ 
corresponding to $\ttt_{\g}$.
The numerator of (\ref{eq:AutGH1}) is mapped by 
this homomorphism $\varphi$ onto the subgroup 
$\text{Aut}_{Q^{\h}}(Q^{\g})\subset \text{Aut}(Q^{\g})$ 
that keep the root lattice $Q^{\h}$ of $\h$ fixed
\beq
   \text{Aut}_{Q^\h}(Q^\g) = \{\omega^\g \in \text{Aut}(Q^\g)\,|\, 
         \omega^\g(Q^\h) =Q^\h\}.
\eeq
On the other hand, the denominator is the inverse image of 
the Weyl group $W(\h)$ of $\h$, {\it i.e.}, 
$\text{Aut}_{\ttt_\g}(\g)\cap \text{Ad}(H)=\varphi^{-1}(W(\h))$.
From the homomorphism theorem, one therefore obtains
the following isomorphism 
\beq
  \text{Aut}_{\ttt_\g}(\g)\cap \text{Aut}_{\h}(\g)
 / \text{Aut}_{\ttt_\g}(\g)\cap \text{Ad}(H)
\cong \text{Aut}_{Q^{\h}}(Q^{\g}) / W(\h).
\eeq
In conclusion, when $\text{rank}\, \g =\text{rank}\, \h$,
one can express the group $\text{Aut}(G/H)$ of induced automorphisms 
in terms of the automorphism group $\text{Aut}(Q^{\g})$ of the root lattice
as follows 
\beq
   \text{Aut}(G/H) \cong \text{Aut}_{Q^{\h}}(Q^{\g}) / W(\h).
\label{eq:AutGHlattice}
\eeq 

The twisted Cardy states corresponding to $\hat{\omega}$ are written 
as~\cite{Ishikawa,IshikawaYamaguchi}
\beq
  \ket{(\tilde{\alpha};\tilde{\beta})}_{\hat{\omega}}
       = \sum_{(\lambda;\mu)\in \hat{\I}^{\hat{\omega}}}
         \frac{\hat{\tilde{S}}_{(\tilde{\alpha};\tilde{\beta})(\lambda;\mu)}}
              {\sqrt{\hat{S}_{(0;0)(\lambda;\mu)}}}
              \dket{(\lambda;\mu)}_{\hat{\omega}},
        \quad~~ (\tilde{\alpha};\tilde{\beta})\in \hat{\tilde{\I}},
\label{eq:twistCardy}
\eeq
where the boundary state coefficients $\hat{\tilde{S}}$
consist of those for the twisted Cardy states 
(\ref{eq:twistedCardy})
associated with the automorphism $\omega^{\g}$ and $\omega^{\h}$,
\beq
    \hat{\tilde{S}}_{(\tilde{\alpha};\tilde{\beta})(\lambda;\mu)}
              = N \tilde{S}^G_{\tilde{\alpha}\lambda}
                  \overline{\tilde{S}^H_{\tilde{\beta}\mu}}.
\label{eq:cosettwisted}
\eeq
For $N$ and other definitions, see below.
Similarly to the case of the WZW models (\ref{eq:twistedIshibashi}), 
the set labeling the twisted 
Ishibashi states is the set of all the representations 
fixed by $\hat{\omega}$
\footnote{
Note that the last equality does not hold in general
since $(\omega^{\g}(\mu);\omega^{\h}(\nu))$ may be equal to $(\mu;\nu)$
due to the field identification.
If this is the case, the construction of the twisted Cardy states 
suffers from the brane identification fixed points~\cite{IshikawaYamaguchi}.}, 
\beq
   \hat{\I}^{\hat{\omega}}= \{(\lambda;\mu) \in \hat{\I}\,|\,
                              \hat{\omega}((\lambda;\mu))=(\lambda;\mu)\}
                          = \{(\lambda;\mu) \in \hat{\I}\,|\,
                              \omega^{\g}(\lambda) =\lambda,~
                              \omega^{\h}(\mu)=\mu \}.
\label{eq:cosettwistspec}
\eeq
As is seen in the definition \eqref{eq:twistCardy}, the labels of 
the twisted Cardy states are
expressed by pairs of those for the $G$ and the $H$ theories.
In the way analogous to the case of the regular Cardy states, 
we need some selection rule and identification of labels
in order to obtain a consistent set of states,
namely, the brane selection rule and 
the brane identification~\cite{Ishikawa,IshikawaTani}.
For the case of regular Cardy states,
both of the selection rule and the identification are described by
the identification current group $\Gid$ 
since the labels of the Ishibashi states and the Cardy states
belong to the same set $\hat{\I}$.
On the other hand, in the case of the twisted Cardy states,
the set labeling the Cardy states is distinct from that 
for the Ishibashi states,
and we need two types of identification groups
$\Gid(\I^{\omega})$ and $\Gid(\tilde{\I})$ to 
define the set labeling the twisted Cardy states.
One can determine these groups once the twisted Cardy states
for the $G$ and the $H$ theories are given~\cite{Ishikawa,IshikawaTani}.
See appendix \ref{sec:twistid} for the definition.
The set of the labels for the twisted Cardy states is then defined as
\beq
\begin{split}
   \hat{\tilde{\I}} &= \{ (\tilde{\alpha};\tilde{\beta})|\,
                \tilde{\alpha}\in \tilde{\I}^G, \tilde{\beta}\in \tilde{\I}^H,
            \tilde{b}_{\tilde{\alpha}}^G(J)=\tilde{b}_{\tilde{\beta}}^H(J'),\,
            \forall (J/J')\in \Gid(\I^{\omega});   \\
            &\hspace{4.8cm}
            (J\tilde{\alpha};J'\tilde{\beta})=(\tilde{\alpha};\tilde{\beta}),\,
            \forall (J/J')\in \Gid(\tilde{\I})\}.
\label{eq:cosettwistedCardy}
\end{split}
\eeq 
The phase $\tilde{b}^G_{\tilde{\alpha}}(J)$ and the action 
$\tilde{\alpha} \mapsto J\tilde{\alpha}$ of the simple currents
on the Cardy states are also defined in appendix \ref{sec:twistid}.
The coefficient $N$ in (\ref{eq:cosettwisted}) is given by the
order of the identification groups, namely,
\beq
   N=|\Gid(\I^{\omega})|=|\Gid(\tilde{\I})|.
\eeq

\section{Boundary conditions in Kazama-Suzuki models}
\label{sec:conditions}
  
\subsection{Kazama-Suzuki models}
\label{sec:KSmodels}
The Kazama-Suzuki models~\cite{KS,KS2} are rational N=2 superconformal 
field theories obtained by applying the coset construction to the 
N=1 super Kac-Moody algebras $\h \subset \g$.
The N=2 superconformal symmetry is obtained when the coset space $G/H$ 
is a K\"{a}hler manifold.
In this subsection, we review some basic facts about these models
and give an argument about the boundary conditions induced 
from those for the current algebras. 

The N=1 super Kac-Moody algebra of $G$ at level $\tilde{k}$
is expressed in terms of the superfields as follows,
\beq
\begin{split}
 & \J^A(z_1,\theta_1)~\J^B(z_2,\theta_2) \sim 
                      \frac{1}{z_{12}}\tilde{k}\, \delta^{A B}
                     +\frac{\theta_{12}}{z_{12}}i\, 
                     {f^{AB}{}_{C}}~\J^{C}(z_2,\theta_2),  \\
 & \hspace{1cm}
     (z_{12} \equiv z_1-z_2-\theta_1 \theta_2, ~~ 
       \theta_{12}\equiv \theta_1-\theta_2).
\label{eq:JJ}
\end{split}
\eeq
We take the orthonormal basis with respect to the Killing metric,
in which the length of the long root is $\sqrt{2}$.
The current $\J^A$ is expressed in terms of the components as 
\beq
   \J^A (z,\theta) = j^A(z) + \theta J^A(z),
\label{eq:supercurrent}
\eeq
where $J^A$ is the bosonic current and $j^A$ is 
its superpartner.
The super stress-energy tensor reads
\beq
 \T_G(z,\theta)= \frac{1}{2\tilde{k}}\Bigl(:D \J^A\, \J_A: 
               + \frac{1}{3\tilde{k}} i :f_{ABC}\J^A \J^B \J^C:\Bigr), 
\label{eq:G}
\eeq
where $D=\partial/\partial \theta + \theta \partial/\partial z$.

Let $H$ ($\h$) be a subgroup (subalgebra) of $G$ ($\g$).
We use the following notations for the indices of
the currents: $A,B,\dotsc$ for $\g$; 
$a,b,\dotsc$ for $\h$; $\bar{a},\bar{b},\dotsc$
for $\g \smallsetminus \h$. 
The super stress-energy tensor of the $G/H$ theory is defined
as the difference of those of the $G$ and the $H$ theories,  
\beq
  \T_{G/H} = \T_{G} - \T_{H}.
\label{eq:cosetEM0}
\eeq
The central charge of this theory is given by
\beq
  c_{G/H} = \frac{3}{2}\left[
             \Bigl(1- \frac{2h^{\vee}_G}{3\tilde{k}}\Bigr)\,\text{dim}\,G
            -\Bigl(1- \frac{2h^{\vee}_H}{3\tilde{k}}\Bigr)\,\text{dim}\,H 
                       \right].
\eeq

In general, the N=1 superconformal algebra enhances to N=2
if and only if there exists a weight one superprimary field $\GG$ 
satisfying the following OPE with itself (see appendix \ref{sec:KSsuper}
for the detail),
\beq
 \GG(z_1,\theta_1)\GG(z_2,\theta_2)
       \sim \frac{1}{z^2_{12}}\,\frac{c}{3}
           +\frac{\theta_{12}}{z_{12}}\,2\, \T(z_2,\theta_2).         
\eeq
For the $G/H$ theory, in addition to this, 
the superprimary field $\GG$ should commute with the $H$ currents $\J^a$.
The most general form of the weight one superfield available 
in the $G/H$ theory is written as
\beq
    \GG_{G/H} = \frac{i}{2\tilde{k}}
               ( \epsilon_A D\J^{A}
                +h_{AB}\J^{A}\J^{B}).
\label{eq:cosetG0}
\eeq
For this $\GG_{G/H}$ together with $\T_{G/H}$ to define
an N=2 SCA, the coefficients $\epsilon_A$ and $h_{AB}$
have to satisfy the following conditions~\cite{KS}
\footnote{
Here we use the Killing metric $g_{AB}=\delta_{AB}$ to
raise and lower the indices.
},
\beq
\begin{split}
   & h^{\bar{b}}{}_{\bar{a}}h^{\bar{a}}{}_{\bar{c}}
          = -\delta^{\bar{b}}{}_{\bar{c}}, 
    \quad  h_{ab}=h_{a\bar{b}}=0,                               \\
   & f^{a \bar{b}}{}_{\bar{d}}h_{\bar{b}\bar{c}}
   - f^{a \bar{b}}{}_{\bar{c}}h_{\bar{b}\bar{d}} =0,                    \\
   & f_{\bar{a}\bar{b}\bar{c}}= 
       h^{\bar{p}}{}_{\bar{a}}h^{\bar{q}}{}_{\bar{b}}f_{\bar{p}\bar{q}\bar{c}}
     + h^{\bar{p}}{}_{\bar{b}}h^{\bar{q}}{}_{\bar{c}}f_{\bar{p}\bar{q}\bar{a}}
     + h^{\bar{p}}{}_{\bar{c}}h^{\bar{q}}{}_{\bar{a}}f_{\bar{p}\bar{q}\bar{b}}, \\
  & \epsilon_{A}= -i f_{A}{}^{\bar{b}\bar{c}}h_{\bar{b}\bar{c}}.
\label{eq:KSconditions0} 
\end{split}
\eeq
The first condition means that $h_{\bar{a}\bar{b}}$ is a
complex structure on the coset space $G/H$, which is invariant 
with respect to $\h$ by the second condition.

The solutions to these conditions are classified in 
\cite{KS2,Schweigert,FSnonHSS}.
We restrict ourselves to the simplest cases, in which
$\text{rank}\,\g = \text{rank}\,\h$ and $\h$ has a single $\uu(1)$ factor.
In these cases, $\h$ takes the form $\h'\oplus \uu(1)$, 
where the Dynkin diagram of $\h'$ is obtained by deleting 
a node from that of $\g$.
We denote this distinguished node of $\g$ by a cross $\times$
and the corresponding simple root and the fundamental weight 
by $\alpha_{\times}$ and $\Lambda_{\times}$, respectively.
The coset we consider is therefore the form of $G/H'\times U(1)$ 
and is specified by $\g$ and the node $\times$ of $\g$.

The conditions \eqref{eq:KSconditions0} become simple 
if $f_{\bar{a}\bar{b}\bar{c}}$ vanish, for which
the corresponding coset space is a hermitian symmetric space (HSS).
The Kazama-Suzuki model based on this coset space is called a HSS model.
In terms of Dynkin diagrams, the HSS models are characterized 
by the condition that a node $\times$ has a unit mark.
\bigskip

We turn to the discussion of boundary conditions in the Kazama-Suzuki models.
Since we are using the superfield formalism, 
we have to specify the boundary condition for the supercoordinate, 
\beq
    \theta - i \eta \tilde{\theta}=0,
\eeq
where $\eta=\pm 1$ depending on the spin structure of the worldsheet.
As we have discussed in section \ref{sec:WZWmutual},
the regular boundary condition for the bosonic current $J^A$
in the charge conjugation modular invariant takes the form 
\beq
    J^A+ \tilde{J}^A=0.
\eeq
The corresponding boundary condition for the supercurrent $\J^A$
is written as 
\beq
    \J^A + i \eta \tilde{\J}^A=0. 
\label{eq:srbc}
\eeq
From eqs.~\eqref{eq:G} and \eqref{eq:cosetG0},
one can show that this induces the following boundary conditions
for the N=2 SCA, 
\beq
\begin{split}
    & \T_{G/H}  - i\eta \tilde{\T}_{G/H}=0,               \\
    & \GG_{G/H}   +\tilde{\GG}_{G/H}  =0,
\end{split}
\eeq
which is written in terms of the components as 
\beq
\begin{split}
  &  T-\tilde{T}=0,                                      \\
  &  G^{\pm}-i\eta \tilde{G}^{\pm} =0,                   \\
  &  J+\tilde{J} =0.
\end{split}
\eeq
Thus the regular boundary condition for $\J^A$ 
in the charge conjugation modular invariant 
yields the B-type boundary condition~\cite{Ooguri} 
for the N=2 SCA. 
On the other hand, as we will show in section \ref{sec:enhance}
(see eq.~\eqref{eq:Atype}),
the regular boundary condition \eqref{eq:dbc}
in the diagonal modular invariant is of the A-type.

\subsection{Automorphisms of the Kazama-Suzuki models}
\label{sec:boudarycond}

As is explained in section \ref{sec:WZWmutual}, 
if there exists an automorphism of the chiral algebra, 
one can use it to twist boundary conditions and obtain 
the corresponding twisted Cardy states compatible with
the regular ones.
The first thing we have to do is therefore the classification 
of automorphisms in the Kazama-Suzuki models, which we consider 
in this subsection.

One can obtain automorphisms of the coset $G/H$ 
from those of $G$, as we have discussed in section \ref{sec:cosetmutual}.
When $\text{rank}\, \g = \text{rank}\, \h$, 
the classification of induced automorphisms is equivalent to
the classification of automorphisms $\omega^{\g}$
of the root lattice $Q^{\g}$ that keeps $Q^{\h}$ fixed
(see eq.(\ref{eq:AutGHlattice})),
\beq
    \omega^{\g} \in \text{Aut}(Q^{\g}), \quad \quad  
    \omega^{\g}(Q^{\h})=Q^{\h}.
\eeq
The latter condition has a simple expression 
for the models we consider.    
As we have mentioned in the previous subsection,
the models we consider have the form $G/H'\times U(1)$,
{\it i.e.}, $\h=\h'\oplus \uu(1)$.
The model in this class is specified by $\g$ and a node $\times$
in the unextended Dynkin diagram of $\g$,
which defines the subalgebra $\h'\subset \g$.
The $\uu(1)$ direction is perpendicular to the root lattice
$Q^{\h'}$ of $\h'$ and is parallel to the fundamental weight $\Lambda_{\times}$
corresponding to the node $\times$.
This is because $(\alpha_i,\Lambda_{\times})=0$ for $i\neq \times$
and $\{ \alpha_i |\, i\neq \times\}$ form the simple roots of $\h'$.
Since $\omega^{\g}$ preserves the angle between any two weights,
the invariance of the $\uu(1)$ direction under $\omega^{\g}$ implies  
that of the root lattice $Q^{\h'}$.
Therefore, $\omega^{\g}$ is an automorphism of $Q^{\h}$ if and only if
it keeps the $\uu(1)$ direction, 
\beq
  \omega^{\g}(\Lambda_{\times}) = \pm \Lambda_{\times}.
\label{eq:branchcond}
\eeq 
In the following, we find the solutions to this equation
(\ref{eq:branchcond})
for any pair of $\g$ and $\Lambda_{\times}$.
\bigskip

The automorphism group $\text{Aut}(Q^{\g})$ is generated by 
the elements of the Weyl group 
$W(\g)$ and the diagram automorphisms of $\g$.
As is seen from (\ref{eq:AutGHlattice}),
$\omega^{\g}\in W(\h)$ gives a trivial automorphism of 
$G/H'\times U(1)$.
Therefore, we have to consider the following two cases,
\bea
  &  \text{(i)}~~~ \omega^{\g} \in W(\g), \quad ~~ \omega^{\g}|_H
                                        :~\text{an outer automorphism of $\h$},
                                                    \nonumber \\
  &  \text{(ii)}~~ \omega^{\g} \notin W(\g).        \nonumber
\eea
We denote by $\text{Aut}(G/H'\times U(1))$ the group of induced 
automorphisms of $G/H'\times U(1)$.
The automorphisms of the case (i) form a subgroup of 
$\text{Aut}(G/H'\times U(1))$, which we call $\text{Aut}_0(G/H'\times U(1))$.
\bigskip
\\
\noindent
\underline{Case (i)}
\smallskip
\\
Let us begin with the case of $\omega^{\g} \in W(\g)$.
As in \eqref{eq:branchcond}, $\omega^{\g}$ has to keep 
$\Lambda_{\times}$ fixed up to sign.
We first consider the case of
$\omega^{\g}(\Lambda_{\times})=\Lambda_{\times}$.
The elements of $W(\g)$ that fix $\Lambda_{\times}$ form a stabilizer group
$S(\Lambda_{\times}) \subset W(\g)$.
One can show that $S(\Lambda_{\times})$ is generated by all the fundamental 
reflections fixing $\Lambda_{\times}$
(see, {\it e.g.}, Proposition 3.12 in \cite{Kac}).
In the present setting, therefore, $S(\Lambda_{\times})$ is given by 
the Weyl group $W(\h')$ of $\h'$,
and $\omega^{\g}(\Lambda_{\times})=\Lambda_{\times}$ means 
$\omega^{\g}\in W(\h') (=W(\h))$.
For $\omega^{\g}$ to be non-trivial, therefore, $\omega^{\g}$
has to map $\Lambda_{\times}$ to $-\Lambda_{\times}$.
This is possible if and only if $\text{{\sf L}}_{\Lambda_{\times}}$
is a real representation of $G$, in which the longest element
$w_0 \in W(\g)$ maps $\Lambda_{\times}$ to $-\Lambda_{\times}$.
$w_0$ is non-trivial in the coset theory, since it flips the
sign of the $\uu(1)$ current.
All the remaining elements that map $\Lambda_{\times}$ to 
$-\Lambda_{\times}$ are obtained by the action of $W(\h')$
and equivalent to $w_0$ as automorphisms of the coset theory.
To summarize, within this class, the  
automorphism group $\text{Aut}_0(G/H'\times U(1))$ takes the
following form,
\beq
  \text{Aut}_0(G/H'\times U(1))=\left\{
                               \begin{array}{ll}
                               \{1,w_0\} \cong \Z_2 
                                 & \quad 
                                   \text{{\sf L}}_{\Lambda_{\times}}
                                   \text{: real},\\ 
                                1& \quad \text{otherwise}. 
                               \end{array}
                               \right.
\eeq 
\bigskip
\\
\noindent
\underline{Case (ii)}   
\smallskip
\\
This type ($\omega^{\g} \notin W(G)$) of automorphisms
contain non-trivial diagram automorphisms, which 
exist only for $\g= A_l$, $D_l$ and $E_6$:
\bea
\text{(a)}~~&\text{charge conjugation $\omega_{c}$ for $A_l$, 
                                $D_{\text{odd}}$ and $E_6$},  \nonumber \\
\text{(b)}~~&\text{chirality flip $\omega_2$ for $D_{\text{even}}$},
                                                              \nonumber \\
\text{(c)}~~&\text{triality $\omega_3$ for $D_{4}$}.          \nonumber
\eea
In general, $\omega^{\g}$ does not keep the root 
lattice $Q^{\h'}$ of $\h'$ invariant.
Thus we have to find an element $w\in W(\g)$ so that
the condition (\ref{eq:branchcond}) is satisfied,
\beq
w\omega^{\g}(\Lambda_{\times})=\pm \Lambda_{\times}.
\label{eq:outercond}
\eeq
It is sufficient to find a particular solution $w$ of this equation;
the others are obtained by the composition with the elements of 
$\text{Aut}_0(G/H'\times U(1))$.
For case (a), it is always possible to find $w$ such that
$w \omega_c (\Lambda_{\times})=-\Lambda_{\times}$ since
$\omega_{c}(\text{{\sf L}}_{\Lambda_{\times}})
=\text{{\sf L}}_{\overline{\Lambda}_{\times}}$
and $\text{{\sf L}}_{\overline{\Lambda}_{\times}}$ includes 
$-\Lambda_{\times}$.
For case (b), $\omega_2$ flips the chirality of spinors,
$\omega_2:~\Lambda_l \leftrightarrow \Lambda_{l-1}$,
where $\Lambda_{l-1}$ and $\Lambda_{l}$ are spinor weights of $D_l$.
Thus $\omega_2(\Lambda_{\times})=\Lambda_{\times}$ if 
$\Lambda_{\times}$ corresponds to a tensor representation.
If $\Lambda_{\times}$ corresponds to the spinors,
there is no solution to eq.~(\ref{eq:outercond}). 
For case (c), the triality $\omega_3$ acts on the 
nodes of $D_4$ as follows:
$\Lambda_1 \rightarrow \Lambda_3 \rightarrow \Lambda_4 
\rightarrow \Lambda_1$, $\Lambda_2 \rightarrow \Lambda_2$.
Thus eq.~(\ref{eq:outercond}) holds for $\Lambda_2$, 
whereas there is no solution
to eq.~(\ref{eq:outercond}) for $\Lambda_{\times}\neq \Lambda_{2}$.
\bigskip

Putting these things together, we obtain the automorphism 
group $\text{Aut}(G/H'\times U(1))$ of the coset theory.
See Table \ref{table:cosetAut} for our result.
\begin{table}[tb]
\begin{center}
\begin{tabular}{|l|l||l|l|}
\hline
~\,$\g$     &  \, $\Lambda_{\times}$
                &  $\text{Aut}_0(G/H'\times U(1))$
                            & $\text{Aut}(G/H'\times U(1))$    \\ \hline\hline
~\,$A_{l}$  &   $\left.
             \begin{array}{l}
              \text{real}  \\
              \text{otherwise}
             \end{array}
             \right.$ 
                &  $\left.
                   \begin{array}{l}
                   \Z_2  \\
                    1
                   \end{array}
                   \right.$
                               & $\left.
                                 \begin{array}{l}
                                 \Z_2\times \Z_2  \\
                                 \Z_2
                                 \end{array}
                                 \right.$                  \\  \hline
~\,$D_{\text{odd}}$
         & $\left.
           \begin{array}{l}
           \text{real(=tensor)}  \\
           \text{spinor}
           \end{array}
           \right.$
               & $\left.
                 \begin{array}{l}
                 \Z_2  \\
                 1
                 \end{array}
                 \right.$    
                    & $\left.
                      \begin{array}{l}
                      \Z_2\times \Z_2 \\
                      \Z_2
                      \end{array}
                      \right.$                             \\ \hline
~\,$D_{\text{even}\neq 4}$
         & $\left.
           \begin{array}{l}
           \text{tensor}  \\
           \text{spinor}  
           \end{array}
           \right.$
              & $\left.
                \begin{array}{l}
                \Z_2 \\
                \Z_2
                \end{array}
                \right.$ 
                    & $\left.
                      \begin{array}{l}
                      \Z_2\times \Z_2 \\
                      \Z_2
                      \end{array}
                      \right.$                     \\ \hline
~\,$D_4$  & $\left.
         \begin{array}{l}
         \Lambda_{2}  \\
         \Lambda_{1}  \\
         \Lambda_{3}, \Lambda_{4} 
         \end{array}
         \right.$
             & $\left.
               \begin{array}{l}
               \Z_2 \\
               \Z_2 \\
               \Z_2
               \end{array}
               \right.$
                  & $\left.
                    \begin{array}{l}
                    S_3 \times \Z_2 \\
                    \Z_2 \times \Z_2 \\
                    \Z_2
                    \end{array}
                    \right.$                      \\ \hline
~\,$E_6$  & $\left.
         \begin{array}{l}
         \text{real} (\Lambda_3,\Lambda_6)  \\
         \text{otherwise}
         \end{array}
         \right.$
            & $\left.
              \begin{array}{l}
              \Z_2  \\
              1
              \end{array}
              \right.$ 
                  & $\left.
                    \begin{array}{l}
                    \Z_2\times \Z_2 \\
                    \Z_2  
                    \end{array}
                    \right.$                   \\ \hline
$\left.
\begin{array}{l}
B_l \\
C_l \\
E_7, E_8 \\
F_4 \\
G_2
\end{array}
\right.$
      &  ~\,any
          &  ~\,$\Z_2$
              &  ~\,$\Z_2$                        \\ \hline     
\end{tabular}
\caption{Automorphism group $\text{Aut}(G/H'\times U(1))$ of the coset 
theories induced from the automorphisms of the $G$ theory.
We specify coset theories by $\g$ and $\Lambda_{\times}$,
the fundamental weight corresponding to the node of $\g$ which is 
not contained in $\h'$.
$\text{Aut}_0(G/H'\times U(1))$ is the subgroup induced from 
the inner automorphism of $\g$.}
\label{table:cosetAut}
\end{center} 
\end{table}
One can see that the automorphism group always contains a $\Z_2$ as a
subgroup, which is nothing but the charge conjugation of 
the coset theory.
In several models, however, the automorphism group
has extra elements other than the charge conjugation,
and hence is larger than $\Z_2$.
We list all the HSS models with extra automorphisms in Table \ref{table:twist}.
The non-HSS models in \cite{FSnonHSS} with extra automorphisms
are also listed.

\begin{table}
\begin{center}
\begin{tabular}{|l|l|}
\hline
\hspace{1.5cm}   models     & \hspace{1.5cm} Dynkin diagrams  \\ \hline\hline
\hspace{1cm}HSS models      &                             \\ \hline
\begin{tabular}{ll}
                                             &  \\
$SU(2n)/SU(n)\times SU(n)\times U(1)$& 
\end{tabular}
                            & \renewcommand{\arraystretch}{0.5}  
                              \begin{tabular}{ll}
                                                                          &\\
                               \includegraphics[width=6cm]{su2msumsum.eps}&
                              \end{tabular}                                 \\
~$SO(2n)/SO(2n-2)\times U(1)$
                            &  \renewcommand{\arraystretch}{0.5}
                               \begin{tabular}{ll}
                                                                         &\\
                              \includegraphics[width=4cm]{sonson2.eps}
                              \end{tabular}
                                                          \\ \hline\hline  
\hspace{1cm} non-HSS models \cite{FSnonHSS}&                \\ \hline
~$SO(8)/SU(2)\times SU(2)\times SU(2)\times U(1)$
                            &  \renewcommand{\arraystretch}{0.4}
                            \begin{tabular}{ll}
                                                                           & \\
                           \includegraphics[width=1.8cm]{so8su2su2su2.eps} &
                              \end{tabular}                                 \\
~$SO(10)/SU(3)\times SU(2)\times SU(2)\times U(1)$
                            &  \renewcommand{\arraystretch}{0.4}  
                              \begin{tabular}{ll}
                                                                           &\\ 
                          \includegraphics[width=2.5cm]{so10su3su2su2.eps} &
                              \end{tabular}                                 \\ 
~$SO(10)/SU(4)\times SU(2)\times U(1)$
                            &  \renewcommand{\arraystretch}{0.4}  
                              \begin{tabular}{ll}
                                                                           & \\
                              \includegraphics[width=2.5cm]{so10su4su2.eps}&
                              \end{tabular}
                                                           \\ \hline
\end{tabular}
\end{center}
\caption{The Kazama-Suzuki models with non-trivial automorphisms 
other than the charge conjugation.
The list covers all the HSS models together with
the non-HSS models studied in \cite{FSnonHSS}.
A cross $\times$ expresses the node $\times$ defining the model.  
The arrows stand for the actions of the extra automorphisms.}
\label{table:twist}
\end{table}

\subsection{Boundary condition and N=2 superconformal symmetry}
\label{sec:enhance}

In the previous subsection, we have classified the
automorphism $\hat{\omega}$ of the Kazama-Suzuki models $G/H'\times U(1)$,
which is induced from the automorphism $\omega^{\g}$ of the $G$ theory.
As we have reviewed in section \ref{sec:KSmodels},
the Kazama-Suzuki models have N=2 superconformal 
symmetry.
It is therefore natural to ask how the automorphisms 
we have found act on the N=2 superconformal algebra.

Let us first consider the action of $\hat{\omega}$
on the N=1 superconformal algebra.
$\omega^{\g}$ keeps the Killing form and the 
structure constants of both of the $G$ and the $H$ theories
($H=H'\times U(1)$).
The super stress-energy tensors $\T_{G}$ and $\T_{H}$
(see (\ref{eq:G})) are therefore invariant under the action of $\omega^{\g}$.
Since the super stress-energy tensor $\T_{G/H}$ (\ref{eq:cosetEM0}) 
of the coset theory is defined as the difference of $\T_G$ and $\T_H$, 
it is also invariant by the action of $\omega^{\g}$,
and hence by the induced automorphism $\hat{\omega}$, 
\beq
    \hat{\omega}(\T_{G/H}) = {\T}_{G/H}.
\eeq

We next turn to the study of the action on the N=2 SCA.
The N=2 SCA consists of the super stress-energy tensor $\T$
and the weight one superprimary field $\GG$ (see appendix \ref{sec:KSsuper}).
For the Kazama-Suzuki models, $\GG$ takes the form given in 
eq.~(\ref{eq:cosetG0}),
\beq
    \GG_{G/H} = \frac{i}{2\tilde{k}}
               ( -i f_A{}^{\bar{b}\bar{c}}h_{\bar{b}\bar{c}}D\J^{A}
                +h_{\bar{a}\bar{b}}\J^{\bar{a}}\J^{\bar{b}}).
\label{eq:cosetG1}
\eeq
Since $\omega^{\g}$ keeps $H'\times U(1)$ invariant, 
the action of $\omega^{\g}$ on the currents can be expressed as follows,
\beq
    \omega^{\g}:~ \J^A \mapsto  \omega^{\g}(\J^A)
                               ={\Omega}^{A}{}_B\J^B,\quad \quad
    {\Omega}^A{}_B = \left(
            \begin{array}{ccc}
            \Omega^{\bar{a}}{}_{\bar{b}}&   0               & 0      \\
                      0                  & {\Omega}^{a}{}_{b}& 0   \\
                      0                  &   0         &\Omega^{0}{}_{0}   
            \end{array}
            \right),     
\label{eq:Omega}
\eeq
where we denote the $\uu(1)$ part by $A=0$, and the indices 
$a,b,\dots$ stand for $\h'$. 
Then the action of $\hat{\omega}$ on $\GG_{G/H}$ takes the form
\beq
\begin{split}
  \hat{\omega}(\GG_{G/H}) 
         &  = \frac{i}{2\tilde{k}}
               \left( -i f_A{}^{\bar{b}\bar{c}}h_{\bar{b}\bar{c}}
                        \Omega^A{}_{A'} D\J^{A'}
                +h_{\bar{a}\bar{b}}\Omega^{\bar{a}}{}_{\bar{a}'}
                                   \Omega^{\bar{b}}{}_{\bar{b}'}
                                   \J^{\bar{a}'}\J^{\bar{b}'}\right)    \\
         &  =\frac{i}{2\tilde{k}}
               \left( -i f_{A'}{}^{\bar{b}'\bar{c}'}h_{\bar{b}\bar{c}}
                          \Omega^{\bar{b}}{}_{\bar{b}'} 
                          \Omega^{\bar{c}}{}_{\bar{c}'}D\J^{A'}
                +h_{\bar{a}\bar{b}}\Omega^{\bar{a}}{}_{\bar{a}'}
                                   \Omega^{\bar{b}}{}_{\bar{b}'}
                                   \J^{\bar{a}'}\J^{\bar{b}'}\right),   
\label{eq:omegaG}
\end{split}
\eeq
where the second equality holds since $\omega^{\g}$ keeps
the Killing form and the structure constants invariant.
Comparing with eq.~(\ref{eq:cosetG1}), one can see that 
$\hat{\omega}(\GG_{G/H})$ is obtained from $\GG_{G/H}$ by replacing 
the complex structure $h_{\bar{a}\bar{b}}$ with its conjugation by $\Omega$,
\beq
h^{\hat{\omega}}_{\bar{a}\bar{b}}\equiv 
               h_{\bar{a}'\bar{b}'}\Omega^{\bar{a}'}{}_{\bar{a}}
                                   \Omega^{\bar{b}'}{}_{\bar{b}}.
\eeq
This $h^{\hat{\omega}}$ can be regarded as a complex structure 
on $G/H$.
Actually, one can check that $h^{\hat{\omega}}$ is also 
a solution to the equations (\ref{eq:KSconditions0}). 
It is therefore natural to ask how many solutions 
the equations (\ref{eq:KSconditions0}) have,
{\it i.e.}, how many complex structures exist on $G/H$.
This problem is studied in appendix \ref{sec:uniqueness},
in which we show that the equations (\ref{eq:KSconditions0})
uniquely determines the complex structure up to sign.
Therefore, one can conclude that 
$h^{\hat{\omega}}$ has to coincide with $h$ up to sign
\beq
   h^{\hat{\omega}}= \pm h, 
\label{eq:omegah}
\eeq
since both of $h$ and $h^{\hat{\omega}}$ satisfy the equations 
(\ref{eq:KSconditions0}).
Substituting this into \eqref{eq:omegaG},
one finds 
\beq
   \hat{\omega}(\GG_{G/H})=\pm \GG_{G/H}.
\label{eq:omegaGpm}
\eeq
As is seen from the OPE of the superfields $\T$ 
and $\GG$, it is clear that the non-trivial automorphism 
of the N=2 SCA that keeps $\T$ invariant is only the sign change 
$\GG \rightarrow -\GG$, which is nothing but the mirror automorphism.
Thus $\hat{\omega}$ acts on the N=2 SCA of the Kazama-Suzuki model 
as the mirror automorphism.

The sign in \eqref{eq:omegaGpm} is correlated with the sign  
appearing in the action \eqref{eq:branchcond} of $\omega^{\g}$ 
on $\Lambda_{\times}$.
In order to see this, let us first recall that 
$\omega^{\g}$ acts on $\J^A$ as $\J^A \mapsto \Omega^A{}_B \J^B$.
From the form \eqref{eq:Omega} of $\Omega$,
it is manifest that $\J^0$ does not mix with the other components,
$\omega^{\g}(\J^0)=\Omega^0{}_0 \J^0 $.
Considering the term proportional to $D \J^0$ in $\GG_{G/H}$ 
\eqref{eq:cosetG1}, one can see that the sign in \eqref{eq:omegaGpm}
is given by $\Omega^0{}_0$.
This is compatible with $\Omega^0{}_0=\pm 1$,
which follows from the fact that $\Omega$ is an orthogonal matrix.
Since $\J^0$ is proportional to $\Lambda_{\times}\cdot H$,
where $H$ are the Cartan currents of $\g$,
the sign of $\Omega^0{}_0$ is the same as that
appearing in \eqref{eq:branchcond}.
We therefore obtain the following relation between the signs
for $\hat{\omega}$ and $\omega^{\g}$,
\beq
   \hat{\omega}(\GG_{G/H})=\pm \GG_{G/H}~~~ \Leftrightarrow~~~
                \omega^{\g}(\Lambda_{\times}) = \pm \Lambda_{\times}.
\label{eq:sign}
\eeq

As we have argued in section \ref{sec:KSmodels},
the regular boundary condition for $\J^A$ 
in the charge conjugation modular invariant is given by eq.~\eqref{eq:srbc},
which yields the B-type boundary condition of the N=2 SCA
\beq
  \GG_{G/H} + \tilde{\GG}_{G/H} =0.
\eeq
Once an automorphism $\omega^{\g}$ is given, 
one can twist boundary conditions for $\J^A$ as follows,
\beq
   \omega^{\g}(\J^A) + i\eta \tilde{\J}^A = 0.
\label{eq:scctbc}
\eeq
The corresponding boundary condition for $\GG_{G/H}$
takes the form
\beq
   \hat{\omega}(\GG_{G/H}) + \tilde{\GG}_{G/H} =0.
\label{eq:Gtbc}
\eeq
Since $\hat{\omega}$ acts as an automorphism of the N=2 SCA 
\eqref{eq:omegaGpm}, one can conclude that 
the twisted boundary condition \eqref{eq:scctbc} 
keeps the N=2 superconformal symmetry,
\beq
    \pm \GG_{G/H} + \tilde{\GG}_{G/H} =0,
\eeq
where the sign is determined according to eq.~\eqref{eq:sign}.

Taking $\omega^{\g}=\omega_c$, the charge conjugation automorphism of $\g$,
we can obtain the regular boundary condition 
for $\J^A$ in the diagonal modular invariant,
\beq
   \omega_c(\J^A) + i \eta \tilde{\J}^A = 0.
\label{eq:sdbc}
\eeq
Since $\omega_c$ maps $\Lambda_{\times}$ to $-\Lambda_{\times}$,
the corresponding boundary condition \eqref{eq:Gtbc} for $\GG_{G/H}$ reads
\beq
   - \GG_{G/H} + \tilde{\GG}_{G/H} =0,
\label{eq:Atype}
\eeq
which is of the A-type.
Similarly to the case of the charge conjugation modular invariant, 
twisted boundary conditions in the diagonal modular invariant
are obtained by using $\omega^{\g}$,
\beq
   \omega^{\g}\omega_c(\J^A) + i\eta \tilde{\J}^A = 0,
\label{eq:sdtbc}
\eeq
which also keeps the N=2 superconformal symmetry.
The boundary condition for $\GG_{G/H}$ in this case is
\beq
   - \hat{\omega}(\GG_{G/H}) + \tilde{\GG}_{G/H} =0.
\eeq
From the relation \eqref{eq:sign}, we find the following correspondence 
\beq
\begin{split}
  & \omega^{\g}(\Lambda_{\times})=\Lambda_{\times}
                        ~~~~~\Leftrightarrow ~~~\text{A-type},  \\
  & \omega^{\g}(\Lambda_{\times})=-\Lambda_{\times}
                        ~~~\Leftrightarrow~~~ \text{B-type}. 
\label{eq:signrule}
\end{split}   
\eeq
Obviously, the trivial element of the automorphism 
group $\text{Aut}(G/H'\times U(1))$ corresponds to
the regular boundary condition, which is of the A-type.
Moreover, the charge conjugation $\omega_c$ is 
always an element of $\text{Aut}(G/H'\times U(1))$,
which yields the B-type boundary condition. 
When $\text{Aut}(G/H'\times U(1))$ contains the
non-trivial elements other than $\omega_c$,
we have additional N=2 boundary conditions,
the type of which is determined by the rule \eqref{eq:signrule}. 

\section{Construction of twisted Cardy states}
\label{sec:examples}
In this section, we give the explicit construction 
of the twisted Cardy states in the Kazama-Suzuki models. 
We restrict ourselves to the case of the diagonal modular 
invariant and hence the regular boundary condition is of the A-type.
\subsection{Bosonic form of the Kazama-Suzuki models}
Since we construct the Cardy states in the bosonic form, 
we first rewrite the supercoset $G/H'\times U(1)$ as 
\beq
   \frac{G_{k}\times SO(2m)_1}
        {{H'}_{I(k+h^{\vee}_G)-h^{\vee}_{H'}} \times U(1)_K},
\eeq
where $2m=\text{dim}\,G/H=\text{dim}\,G/H' -1$,
$I$ is the Dynkin index of the embedding $\h'$ into $\g$  
and the level $K$ of $U(1)$ is model dependent.
The $SO(2m)$ part consists of the 
$2m$ free fermions $j^{\bar{a}}$.
The level $k$ is related to the level $\tilde{k}$ of the super Kac-Moody 
algebra \eqref{eq:JJ} as $k=\tilde{k}-h^{\vee}_G$.
The spectrum $\hat{{\cal I}}$ of this model 
takes the form 
\beq
 \hat{{\cal I}}
          =\{\widehat{\Lambda}\equiv
            (\Lambda, \tilde{\Lambda};~ \lambda, \sigma)|
                  ~\text{selection rule},
                  ~\text{field identification}\} ,
\label{eq:KSprimary}
\eeq
where each entry of $\widehat{\Lambda}$ stands for the integrable 
representation of the constituent theories,
\beq
\begin{split}
  &\Lambda           \in   P_+^k (\g),                 \quad\quad\quad~~~~
  \tilde{\Lambda}   \in   P_+^1(\s\oo (2m))=\{o,v,s,c\},           \\    
  &\lambda           \in   P_+^{I(k+h^{\vee}_G)-h^{\vee}_{H'}}(\h'), \quad
  \sigma            =   0,1,\dots, K-1.          
\label{eq:GHprimary}
\end{split}
\eeq
In the $SO(2m)$ part, the labels $o$, $v$, $s$ and $c$ 
express the vacuum, vector, spinor and cospinor representations, respectively.
The explicit forms of the selection rule
and the field identification are given in appendix \ref{sec:selectid}.
The examples considered in this section have no fixed points
and the matrix $\hat{S}$ (\ref{eq:cosetSmatrix}) of the coset theory
is given by 
\beq
   \hat{S}_{\widehat{\Lambda}\widehat{\Lambda}'} 
           = N_0~
             S^{G_k}_{\Lambda \Lambda'}
             S^{SO(2m)_1}
              _{\tilde{\Lambda} {\tilde{\Lambda}}'}
             \overline{S^{{H'}_{I(k+h^{\vee}_G)-h^{\vee}_{H'}}}
                        _{\lambda \lambda'}
                       S^{U(1)_{K}}_{\sigma \sigma'}},
\eeq
where $N_0$ is the order of the field identification group. 

To construct twisted Cardy states, we have to clarify 
how the automorphism $\omega^{\g}$ acts on the $SO(2m)$ part.
Since $j^{\bar{a}}$ is the component of the supercurrent  
$\J^{\bar{a}}$ \eqref{eq:supercurrent}, 
$\omega^{\g}$ acts on the free fermions $j^{\bar{a}}$ in the same way 
as $\J^{\bar{a}}$,
\beq
  j^{\bar{a}} \mapsto \omega^{\g}(j^{\bar{a}})
         =\Omega^{\bar{a}}{}_{\bar{b}} j^{\bar{b}}, 
\eeq
where we used \eqref{eq:Omega}.
Let $\bar{\Omega}$ be a $2m \times 2m$ 
orthogonal matrix $(\Omega^{\bar{a}}{}_{\bar{b}})$.
If $|\bar{\Omega}|=1$, $\bar{\Omega}$ is an element of $SO(2m)$
and $\omega^{\g}$ acts on the $SO(2m)$ part as an inner automorphism,
whereas, if $|\bar{\Omega}|=-1$, $\omega^{\g}$ acts as the outer automorphism
$s \leftrightarrow c$.

Whether the action of $\omega^{\g}$ on the $SO(2m)$ part is 
inner or outer can be seen from its action on $\Lambda_{\times}$.
Let $\bar{\Delta}$ be the roots in $\g \smallsetminus (\h' \oplus \uu(1))$.
This set $\bar{\Delta}$ are decomposed into two sets
$\bar{\Delta}_{\pm}$ according to the sign of the $U(1)$ charge.
Each set contains $m$ elements.
If $\omega^{\g}(\Lambda_{\times})=\Lambda_{\times}$,
$\omega^{\g}$ maps $\bar{\Delta}_{\pm}$ to itself
and $\bar{\Omega}$ has the block diagonal form 
$\bar{\Omega}=O \oplus O$ on $\bar{\Delta}_+ \oplus \bar{\Delta}_-$,
where $O$ is an orthogonal matrix.
Therefore, $|\bar{\Omega}|=|O|^2=1$ and $\omega^{\g}$ acts 
on the $SO(2m)$ part as an inner automorphism.
On the other hand, if $\omega^{\g}(\Lambda_{\times})=-\Lambda_{\times}$,
$\omega^{\g}$ exchanges $\bar{\Delta}_{+}$ and $\bar{\Delta}_-$.
It is sufficient to consider the case that every positive root is mapped 
to its negative and vice versa, since other cases are obtained 
by the composition with the elements of $SO(2m)$.
In this case, $|\bar{\Omega}|=(-1)^m$, and hence 
$\omega^{\g}$ acts as an inner automorphism for even $m$ and 
as an outer automorphism for odd $m$.

\subsection{Examples}
\label{sec:example}
\subsubsection{$SU(4)/SU(2)\times SU(2) \times U(1)$}
\label{sec:example1}

This model is one of the models based on the complex Grassmannian
manifold $SU(m+n)/SU(m)\times SU(n)\times U(1)$,
\beq
 \frac{SU(m+n)_k \times SO(2mn)_1}
      {SU(m)_{n+k}\times SU(n)_{m+k}\times U(1)_{mn(m+n)(m+n+k)}},
\label{eq:mnk}
\eeq 
with the central charge
\beq
 c=\frac{3m n k}{m+n+k}.
\label{eq:c}
\eeq
We consider the case $(m,n,k)=(2,2,1)$, {\it i.e.}, the model
\beq
   SU(4)_1\times SO(8)_1/SU(2)_3\times SU(2)_3 \times U(1)_{80}.
\eeq
The primary fields of this model are labeled by the following 
representations (see eq.~\eqref{eq:GHprimary}),
\beq
\begin{split}
\Lambda &\in P_+^{1}({\mathfrak s}\uu(4))
         =\{(0,0,0),(1,0,0),(0,1,0),(0,0,1)\},                \\ 
\lambda &\in P_+^{3}({\mathfrak s}\uu(2))\times 
             P_+^{3}({\mathfrak s}\uu(2))
         =\{(\lambda^{(1)},\lambda^{(2)})\,|\, \lambda^{(j)}=0,1,2,3\}.
\end{split}
\eeq
Here each representations are expressed by the Dynkin labels.
The identification current group is generated by 
(see (\ref{eq:suidcurrent}))
\beq
\begin{split}
  J_{(1)} &= (J,1/\,(J',1),10),               \\
  J_{(2)} &= (J,1/\,(1,J'),-10),
\end{split}
\eeq
where $J$ and $J'$ are the generators of the simple current group of 
$SU(4)_1$ and $SU(2)_3$, respectively, while
$\pm 10$ represents the shift of the $\uu (1)$ charge 
$\sigma\mapsto \sigma \pm 10$.
From the relations $J^4=1$ and ${J'}^2=1$, one obtains
\beq
   \Gid \cong \Z_2 \times \Z_8, \quad ~~N_0=|\Gid| = 16.
\eeq
Since this model has no field identification fixed points,
the number of the primary fields in this model is calculated as 
\beq
\begin{split}
|\hat{\I}|&=\frac{|P_+^1(\s \uu (4))|\times |P_+^1(\s \oo(8))|
               \times |P_+^3 (\s\uu(2))\times P_+^3 (\s\uu(2))|
               \times |P_+^{80}(\uu (1))|}
                {|\Gid|\times |\Gid|}                                \\
          &=\frac{4\times 4 \times (4\times 4) \times 80}{16\times16}=80,
\end{split}
\eeq
where the factors in the denominator correspond to
the selection rule and the field identification.
The explicit form of $\hat{\I}$ reads
\bea
  \hat{\cal I}&=\{ ((0,0,0),\tilde{\Lambda};\,(0,0),8j),
                   ((0,0,0),\tilde{\Lambda};\,(1,1),4+8j),  \nonumber \\
       &\hspace{0.8cm}((0,0,0),\tilde{\Lambda};\,(0,2),8j),    
                    ((0,0,0),\tilde{\Lambda};\,(2,0),8j) |\,
                      \,\tilde{\Lambda}\in\{o,v,s,c\};~j =0,1,\dots,4\}.
\label{eq:su4su2su2primary}
\eea 
The modular transformation matrix takes the form
\beq
   \hat{S}_{\hat{\Lambda} \hat{\Lambda}'}
    = 16\, S^{SU(4)_1}_{\Lambda_0 \Lambda_0}
         S^{SO(8)_1}_{\tilde{\Lambda} {\tilde{\Lambda}}'}
         \overline{S^{SU(2)_3}_{\lambda^{(1)} \lambda^{(1)'}}
         S^{SU(2)_3}_{\lambda^{(2)} \lambda^{(2)'}}
         S^{U(1)_{80}}_{\sigma \sigma'}}.
\eeq
One can construct the corresponding 80 regular Cardy states
in the diagonal modular invariant, which are of the A-type,
in the standard manner.

As is shown in Table \ref{table:twist}, 
the automorphism group $\text{Aut}(SU(4)/SU(2)\times SU(2)\times U(1))$
is $\Z_2\times \Z_2$ and contains a non-trivial element 
that fixes $\Lambda_{\times}=\Lambda_2$.
We denote this automorphism by $\hat{\omega}_o$.
The corresponding automorphism $\omega_o$ of $\s\uu (4)$ acts 
as is indicated  by the arrow in Table \ref{table:twist}. 
One can see that $\omega_o$ acts on $\s \uu(2)\oplus \s\uu(2)$ 
as the permutation $\pi$ of two factors.
From the argument in the previous subsection,
$\omega_o$ acts on $\s \oo (8)$ as an inner automorphism,
and the action on the $\uu (1)$ part is trivial,
since $\omega_o(\Lambda_{\times})=\Lambda_{\times}$.
To summarize, the automorphism $\hat{\omega}_o$ acts in each sector
of the coset theory as follows,
\beq
  \hat{\omega}_o = (\omega_o,1;\, \pi,1).
\label{eq:su4su2su2lift}
\eeq

As we have mentioned above, the regular boundary condition 
in the diagonal modular invariant is of the A-type.
Since $\omega_o$ fixes $\Lambda_{\times}$, 
the boundary condition twisted by $\hat{\omega}_o$
is also of the A-type (see \eqref{eq:signrule}).
The corresponding twisted Cardy states are constructed following to
the procedure reviewed in section \ref{sec:cosetmutual}.
The boundary state coefficients $\tilde{S}^G$ for the $G$ theory 
is the tensor product of $\tilde{S}^{SU(4)_1}$ and $S^{SO(8)_1}$
(see section \ref{sec:WZWmutual} for our notation). 
Similarly, $\tilde{S}^H$ is the tensor product of 
$\tilde{S}^{SU(2)_3\times SU(2)_3}=S^{SU(2)_3}$ and $S^{U(1)_{80}}$.
From the formula in appendix \ref{sec:twistid}, we find that 
$|\Gid(\I^{\omega})|=|\Gid(\tilde{\I})|=8$.
Putting these things together, the number of the 
twisted Cardy states is calculated as follows,
\beq
\begin{split}
   |\hat{\tilde{\I}}|
  &= \frac{|P_+^1(A_3^{(2)})|\times|P_+^1(\s\oo (8))| \times 
   |P_+^3(\s\uu(2))|\times |P_+^{80}(\uu (1))|}
   {|\Gid(\I^{\omega})|\times |\Gid(\tilde{\I})|}           \\
  &= \frac{2\times 4\times 4\times 80}{8\times8}=40.
\end{split}
\eeq
We give the explicit form of the twisted Cardy states 
in appendix \ref{sec:suexplicit}.

We obtain 40 A-type twisted Cardy states besides 80 regular ones,
yielding 120 A-type Cardy states in total.
These 120 states have the following natural interpretation.
As is seen from the value of the central charge $c=12/5$, 
the Grassmannian model $(2,2,1)$ is equivalent to one of 
the N=2 minimal models, 
\beq
   \frac{SU(2)_8\times SO(2)_1}{U(1)_{20}}.
\eeq
More precisely, we should take the D-type modular 
invariant in the $SU(2)_8$ theory for the equivalence.
The number of the primary fields in this model is
$
  6\times 4 \times 20 /2/2=120,
$
where the first factor stands for
the number of the primary fields in the $\text{D}_6$ 
modular invariant of $SU(2)_8$.
Correspondingly, by taking the D-type boundary coefficients 
for the $SU(2)_8$ part,
we can construct 120 Cardy states in this minimal model,
which satisfy the N=2 A-type boundary condition.
This is the same number as we have obtained for the
Kazama-Suzuki model $SU(4)/SU(2)\times SU(2)\times U(1)$ 
at level one.
Actually, from the comparison of the boundary state coefficients,
one can identify 80 of the 120 Cardy states in the minimal model 
with the regular ones in the Kazama-Suzuki model,
and 40 of them with the twisted Cardy states we have obtained above.
This result shows that 40 twisted Cardy states are compatible with
80 regular ones in the diagonal modular invariant 
of the Kazama-Suzuki model,
since these two sets are combined into one NIM-rep of the minimal model. 

\subsubsection{$SO(8) /SO(6) \times U(1)$}
\label{sec:example2}

The next example is the first non-trivial model in 
the series of
\beq
\frac{SO(2n)_k\times SO(4(n-1))_1}{SO(2n-2)_{k+2}\times U(1)_{4(k+2(n-1))}}. 
\label{eq:SO}
\eeq
We consider the case of $n=4,k=1$.
The primary fields of this model are labeled by the following representations
\beq
\begin{split}
 & \Lambda \in P_+^1(\s\oo(8))
            = \{o,v,s,c\},            \\     
 & \lambda \in P_+^3({\mathfrak s}\oo(6))
            = \{(\lambda_1,\lambda_2,\lambda_3)|\, 
                 \lambda_1+\lambda_2+\lambda_3 \leq 3\}.
\end{split}
\eeq  
Here $\lambda_2$ ($\lambda_3$) corresponds to the
cospinor (spinor) node of $\s\oo (6)$.
The identification group of this model is given by 
(see appendix \ref{sec:soselectid})
\beq
 G_{\text{id}}\cong \Z_2 \times \Z_4.
\eeq
Since this model has no identification fixed points,
the number of the primary fields is calculated as
\beq
\begin{split}
  |\hat{{\cal I}}| &= \frac{|P_+^1(\s \oo (8))|\times |P_+^1(\s \oo (12))|
                        \times |P_+^3(\s \oo (6))|\times |P_+^{28}(\uu (1))|}
                          {|\Gid|\times |\Gid|}                \\
                   & = \frac{4\times 4 \times 20 \times 28}{8 \times 8}
                     =140.
\end{split}
\eeq
The explicit form of $\hat{{\cal I}}$ reads
\bea
   \hat{{\cal I}}&=\{(o,\tilde{\Lambda}_{\text{NS}}\,;\,
                    (\lambda_1,\lambda_2,\lambda_3),4j),
                   (o,\tilde{\Lambda}_{\text{R}}\,;\,
                    (\lambda_1,\lambda_2,\lambda_3),4j+2) \,
                 |\,\tilde{\Lambda}_{\text{NS}}\in \{o,v\},
                    \tilde{\Lambda}_{\text{R}}\in \{s,c\}; \nonumber \\
                 & \hspace{1cm}\,
                (\lambda_1,\lambda_2,\lambda_3) 
            \in \{(0,0,0),(2,0,0),(0,1,1),(1,2,0),(1,0,2)\};~j=0,1,\dots,6\}.
\eea
We can construct corresponding 140 regular Cardy states 
in the diagonal modular invariant, which are of the A-type.

This models also has an extra automorphism that fixes
$\Lambda_{\times}= \Lambda_1$ (see Table \ref{table:twist}).
We denote this automorphism by $\hat{\omega}_2$,
the action of which on each sector reads
\beq
   \hat{\omega}_2 = (\omega_2,1;\omega'_c,1).
\label{eq:solift}
\eeq 
Here $\omega_2$ is the order two outer automorphism of $\s \oo(8)$
and $\omega'_c$ is the charge conjugation of $\s \oo(6)$.

We can construct the corresponding twisted Cardy states 
in the way parallel to the first example.
We find that $|\Gid(\I^{\omega})|=|\Gid(\tilde{\I})|=4$
and the number of the twisted Cardy states is calculated as
\beq
\begin{split}
   |\hat{\tilde{\I}}|&=\frac{|P_+^1(D_4^{(2)})|
                      \times|P_+^1(\s\oo (12))|
                      \times|P_+^3(D_3^{(2)})|
                      \times|P_+^{28}(\uu(1))|}
                           {|\Gid(\I^{\omega})|\times|\Gid(\tilde{\I})| }\\
                     &= \frac{2 \times 4 \times 6 \times 28}{4\times 4}
                      = 84.   
\end{split}
\eeq 
The explicit form of these twisted Cardy states are given in 
appendix \ref{sec:soexplicit}.

We obtain 84 A-type twisted Cardy states in addition to 
140 regular ones, yielding 224 A-type Cardy states as a whole.
Similarly to the previous example, these states are 
interpreted in terms of the N=2 minimal model,
which is the $k=12$ minimal model with the $\text{D}_8$ modular invariant
this time. 
In this model, there exist $8\times 4\times 28/2/2 = 224$ 
Cardy states of the A-type, and we again see the complete match
of the numbers of the Cardy states for both models.

\subsection{Action of the level-rank duality}
\label{sec:exceptional}

In this subsection, we consider the action of the level-rank
duality\footnote{%
For the level-rank duality in the WZW models, see \cite{KN,NS0}.}
\cite{KS,Gepner0,LVW,Schellekens,FS3,NS} on the twisted Cardy states 
we have obtained.
Some of the Kazama-Suzuki models are related with
each other by exchanging the level and the rank of 
the WZW models.
For example, the Grassmannian models $(m,n,k)$ are invariant 
with respect to the permutation of $m$, $n$ and $k$.
Namely, for $m=n$, the following identity holds,
\beq
  \frac{SU(2m)_k \times SO(2m^2)_1}
       {SU(m)_{k+m}\times SU(m)_{k+m}\times U(1)_{2m^3(k+2m)}}
  =
  \frac{SU(k+m)_m \times SO(2k m)_1}
       {SU(k)_{2m}\times SU(m)_{k+m}\times U(1)_{k m (k+m)(k+2m)}}. 
\label{eq:lrduality}
\eeq
The primary fields, and hence the regular Cardy states, 
for both models are identified by this duality map.
As we have found in section \ref{sec:boudarycond},
the automorphism group $\text{Aut}(SU(2m)/SU(m)\times SU(m)\times U(1))$
is $\Z_2 \times \Z_2$ and one can construct A-type twisted Cardy states 
corresponding to non-trivial automorphism other than the charge conjugation.
On the other hand, in the right hand side of \eqref{eq:lrduality},
$\text{Aut}(SU(k+m)/SU(k)\times SU(m)\times U(1))=\Z_2$ except for $k=m$,
and hence twisted Cardy states cannot be constructed 
in the way similar to the examples we have given.
Then, it is natural to ask what kind of states in the right hand side 
correspond to the twisted Cardy states in the left hand side
via the level-rank duality.

In order to answer to this question, we study the action of
the level-rank duality on the twisted Cardy states 
in $SU(4)/SU(2)\times SU(2)\times U(1)$ constructed 
in section \ref{sec:example1}.
According to eq.~\eqref{eq:lrduality}, this model is dual to the model
\beq
  \frac{SU(3)_2 \times SO(4)_1}{SU(2)_3 \times U(1)_{30}},
\eeq 
for which the automorphism group is $\Z_2$
(see Table \ref{table:cosetAut}) 
and we cannot construct A-type twisted Cardy states in 
the diagonal modular invariant using the method of the
induced automorphism.
 
A primary field in this model is labeled by the following 
representations
\beq
\begin{split}
 &\Lambda \in P_+^2({\mathfrak s}\uu(3))
                         =\{(0,0),(2,0),(0,2),(1,1),(0,1),(1,0)\},
                                                             \\
 &\lambda \in P_+^3({\mathfrak s}\uu(2))=\{0,1,2,3\}.
\end{split}
\eeq
The identification current group is generated by 
\beq
            (J,\tilde{J}_v/J',5),
\eeq
and  $\Gid \cong \Z_6$.
Here $J$ and $J'$ are the generators of the simple current group
for $\s \uu(3)_2$ and $\s \uu(2)_3$, respectively.
$\tilde{J}_v$ is the vector simple current for $\s \oo (4)_1$
and $5$ means the shift of the $\uu(1)$ charge by five.
The number of the primary fields is $6\times 4 \times 4 \times 30 /6/6=80$,
which is equal to that for the model $SU(4)/SU(2)\times SU(2)\times U(1)$
at level one and is consistent with the level-rank duality.
The set $\hat{\I}$ of the primary fields reads
\bea
  \hat{{\cal I}}&=\{ ((0,0),o\,;\,\lambda,6j+3\lambda),
                    ((0,0),s\,;\,\lambda,6j+3+3\lambda),   \nonumber  \\
    & \hspace{0.7cm}((1,1),o\,;\,\lambda,6j+3\lambda),
                    ((1,1),s\,;\,\lambda,6j+3+3\lambda)\,|\,\lambda=0,1,2,3;\,
                                                          j= 0,1,\dots,4 \}.
\eea
The modular transformation matrix takes the form
\beq
   \hat{S} = 6\, S^{SU(3)_2}S^{SO(4)_1}\overline{S^{SU(2)_3}S^{U(1)_{30}}}.
\eeq
From this $\hat{S}$, one can construct  
80 A-type regular Cardy states, which are identified with 
80 regular Cardy states in $SU(4)/SU(2)\times SU(2)\times U(1)$.

We have seen in section \ref{sec:example1} that there are 
40 twisted Cardy states in $SU(4)/SU(2)\times SU(2)\times U(1)$
at level one besides 80 regular ones.
Upon the level-rank duality, these 40 states should be mapped 
to 40 states in $SU(3)/SU(2)\times U(1)$ at level two,
which are compatible with the regular ones.
In $SU(3)/SU(2)\times U(1)$, these 40 states have to
form a NIM-rep of the fusion algebra. 
One can easily see that the resulting NIM-rep cannot be 
constructed by the method of the induced automorphisms
\footnote{
The existence of these states has been reported in \cite{LW}
based on the comparison of the spectrum with that of the minimal model.},
since the automorphism originated from $\s \uu (3)$ gives
no A-type NIM-rep other than the regular 
one (see Table \ref{table:cosetAut}).
Therefore, we have to consider other possibilities in order to 
explain these states within $SU(3)/SU(2)\times U(1)$.
A systematic procedure to construct NIM-reps in coset theories has 
been given in \cite{IshikawaTani}.
In the case of $SU(3)_2\times SO(4)_1/SU(2)_3\times U(1)_{30}$, 
one can apply the method of \cite{IshikawaTani} to obtain a NIM-rep
based on a conformal embedding,
\beq
  {\mathfrak s}\uu(6)_1 \supset {\mathfrak s}\uu(3)_2 
                                \oplus {\mathfrak s}\uu(2)_3.
\label{eq:embedding}
\eeq
As we will see below, this NIM-rep is fourty dimensional,
and precisely coincides with the NIM-rep corresponding to  
the twisted Cardy states in 
$SU(4)/SU(2)\times SU(2)\times U(1)$.\footnote{
The relation between level-rank dualities and conformal embeddings 
has been observed before in the context of the bulk theory.
See {\it e.g.} \cite{Altschuler}.}

\begin{figure}[tb]
\begin{center}
\includegraphics[width=7.5cm]{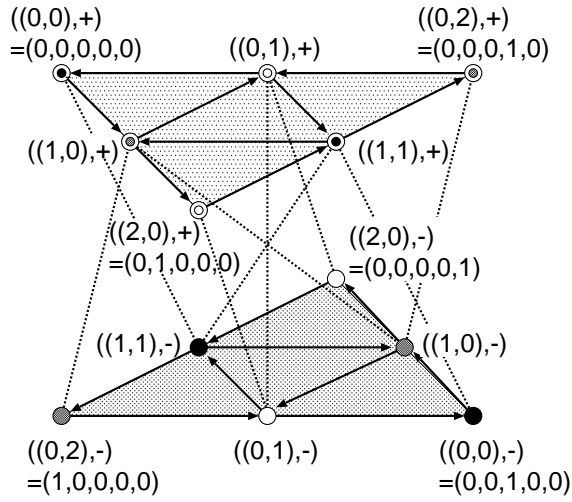}
\end{center}
\caption{The NIM-rep graph of the $SU(3)_2 \times SU(2)_3$ WZW model
from the conformal embedding ${\mathfrak s}\uu(3)_2\oplus {\mathfrak s}\uu(2)_3
\subset {\mathfrak s}\uu(6)_1$. Six of these states are originated from the
regular ones in the $SU(6)_1$ theory. 
The subgraph in the dashed (solid) lines 
corresponds to $n_{((0,0),1)}$ ($n_{((1,0),0)}$).
For the definition of the NIM-rep graph, see {\it e.g.}\cite{IshikawaTani}.
The value $\tilde{b}(J,1)$ is expressed 
as black ($1$), gray ($\omega$) 
or white ($\omega^2$) with $\omega=e^{2 \pi i/3}$, whereas
$\tilde{b}(1,J')$ is expressed as circled ($+1$) or 
un-circled ($-1$).
The action of the simple current of $SU(3)_2$ ($SU(2)_3$) 
is $2 \pi/3$ rotations (the reflection about the center of this diagram).}
\label{fig:su3su2su6}
\end{figure}

The construction of a NIM-rep based on a conformal 
embedding starts from finding the branching of the representations,
\beq
\begin{split}
   (0,0,0,0,0) & \mapsto ((0,0),0) \oplus ((1,1),2),     \\
   (1,0,0,0,0) & \mapsto ((1,0),1) \oplus ((0,2),3),     \\
   (0,1,0,0,0) & \mapsto ((2,0),0) \oplus ((0,1),2),     \\
   (0,0,1,0,0) & \mapsto ((1,1),1) \oplus ((0,0),3),     \\
   (0,0,0,1,0) & \mapsto ((0,2),0) \oplus ((1,0),2),     \\
   (0,0,0,0,1) & \mapsto ((0,1),1) \oplus ((2,0),3),                    
\end{split}
\label{eq:branching}
\eeq
where the left hand sides are the elements of $P_+^1({\mathfrak s}\uu(6))$.
We can construct Cardy states in the $SU(3)_2\times SU(2)_3$ theory
based on this branching.
The general procedure has been given in 
section 2.3.2 of \cite{IshikawaTani} (see also \cite{AOS}).
Since there appear twelve representations  
in the branching \eqref{eq:branching}, we have twelve Ishibashi states,
which we label by the set 
\bea
  {\cal E}^e &=\{((0,0),0),((1,0),2),((2,0),0),
                 ((1,1),2),((0,2),0),((0,1),2),              \nonumber \\
           &~~~~~((0,0),3),((1,0),1),((2,0),3),
                 ((1,1),1),((0,2),3),((0,1),1)  \}.
\label{eq:specsu3su2su6}
\eea
We first obtain six Cardy states by 
reinterpreting six regular states in the $SU(6)_1$ theory.
Next, by the fusion in $SU(3)_2$ and $SU(2)_3$,
we obtain six additional states, and hence 12 Cardy states 
in total which we label by the set 
\bea
  {\cal V}^e &=\{((0,0),+),((1,0),+),((2,0),+),
                 ((1,1),+),((0,2),+),((0,1),+),              \nonumber \\
           &~~~~~((0,0),-),((1,0),-),((2,0),-),
                 ((1,1),-),((0,2),-),((0,1),-)  \}.
\label{eq:branesu3su2su6}
\eea
See Fig.\ref{fig:su3su2su6} for our labeling of the states
and the corresponding NIM-rep graph.

By using these Cardy states in the $SU(3)_2\times SU(2)_3$ theory,
we can construct Cardy states in the coset theory
$SU(3)_2 \times SO(4)_1/SU(2)_3 \times U(1)_{30}$.
In the same way as the case of the induced automorphisms
explained in section \ref{sec:cosetmutual},
we need some selection rule and identification of 
the Cardy states in order to obtain a NIM-rep
in the coset theory.
One can show that the selection rule and the identification 
have the same order as the field identification group, namely,
\beq
  |\Gid({\cal E})|= |\Gid({\cal V})| = 6.
\eeq  
Therefore, the number of the resulting Cardy states is calculated as 
\beq
\begin{split}
   |\hat{\V}^e| &=\frac{|\V^e|
                   \times |P_+^1(\s \oo (4))|
                   \times |P_+^{30}(\uu (1))|}
                     {|\Gid(\E)|\times |\Gid(\V)|}          \\
                &=\frac{12\times 4 \times 30}{6 \times 6}
                 = 40.
\end{split}
\eeq
This number coincides exactly with that of the twisted Cardy states
in $SU(4)/SU(2)\times SU(2)\times U(1)$,
which suggests that this NIM-rep expresses 
the twisted Cardy states we have obtained.
In fact, we can check that the coefficients of the 
twisted Cardy states are precisely reproduced from
this NIM-rep.
The explicit form of the NIM-rep reads
\bea
  \hat{\psi}^e &= 6 \, \psi^e S^{SO(4)_1}\overline{S^{SU(1)_{30}}}, 
                                                                 \nonumber  \\
  \hat{{\cal E}}^e                                
             &=  \{ ((0,0),o;0,6j), ((0,0),o;3,6j+3),
                  ((0,0),s;0,6j+3),((0,0),s;3,6j),           \nonumber  \\
             &
                 \quad~~~ ((1,1),o;1,6j+3),((1,1),o;2,6j),
                  ((1,1),s;1,6j),((1,1),s;2,6j+3)            \nonumber \\
             &     \hspace{10cm}        |\, j=0,1,\dots,4\},
                                                 \nonumber \\
 \hat{{\cal V}}^e &= \{ ((0,0),o;+,6j),((0,0),o;-,6j+3),
                  ((0,0),s;+,6j+3),((0,0),s;-,6j),           \nonumber  \\
             & 
                \quad~~~  ((1,1),o;-,6j+3),((1,1),o;+,6j),
                  ((1,1),s;-,6j),((1,1),s;+,6j+3)\,         \nonumber \\
             &     \hspace{10cm}        |\,j=0,1,\dots,4\},
                                                 \nonumber \\
  \psi^e   & = \frac{1}{\sqrt{2}}
          \left(
          \begin{array}{cc}
           S^{SU(3)_2} & S^{SU(3)_2}  \\
           S^{SU(3)_2} & -S^{SU(3)_2}
          \end{array}
          \right).           
\eea
Here $\psi^e$ is the boundary state coefficients for the NIM-rep
based on the conformal embedding \eqref{eq:embedding} and  
the rows and columns of $\psi^e$ are ordered as 
in (\ref{eq:specsu3su2su6}) and \eqref{eq:branesu3su2su6}.
$S^{SU(3)_2}$ is the $6 \times 6$ modular transformation
matrix whose rows and columns are the $SU(3)_2$ part of
\eqref{eq:specsu3su2su6} and \eqref{eq:branesu3su2su6}.

To summarize, the twisted Cardy states of $SU(4)/SU(2)\times SU(2)\times U(1)$
at level one are mapped via the level-rank duality to those expressed by
the NIM-rep based on the conformal embedding 
$\s \uu (3)_2  \oplus \s \uu (2)_3 \subset \s\uu(6)_1$.
This result shows that the Cardy states compatible with the regular ones 
are not limited to those obtained from the induced automorphisms.
Since the automorphism group of the chiral algebra is considered to be
not affected by the level-rank duality, this result suggests
that the automorphism group of the coset 
theory $G/H$ is in general larger than that induced from 
those of $G$.

\section{Summary and Discussion}
\label{sec:summary}

In this paper, we have studied Cardy states in the 
Kazama-Suzuki models $G/H'\times U(1)$, which satisfy 
the boundary conditions twisted by the automorphisms of the 
coset theory.
We have classified automorphisms of $G/H'\times U(1)$ induced from 
those of the $G$ theory.
The automorphism group of the Kazama-Suzuki models
contains at least a $\Z_2$ as a subgroup which corresponds
to the charge conjugation.
We have found that in several models the automorphism group
contains non-trivial elements other than the charge conjugation
and can be larger than $\Z_2$.
Based on the general procedure to construct the Cardy states 
in coset theories, we have given the explicit form of the twisted Cardy states
corresponding to non-trivial automorphisms of the Kazama-Suzuki models.
We have shown that the resulting twisted Cardy states 
preserve the N=2 superconformal algebra, {\it i.e.}, either of the A-type 
or of the B-type.
As an illustrative example of our construction, 
we have given a detailed study for two HSS models:
$SU(4)/SU(2)\times SU(2)\times U(1)$ and $SO(8)/SO(6)\times U(1)$
both at level one, which have the description as the N=2 minimal models.
We have compared our results with those for the minimal models
and have shown that the twisted Cardy states together with the regular ones
reproduce all the Cardy states for the minimal models.
The action of the level-rank duality on the twisted Cardy states
has been studied for the simplest case, $SU(4)/SU(2)\times SU(2)\times U(1)$
at level one, which is equivalent to $SU(3)/SU(2)\times U(1)$ at level two.
We have shown that the level-rank duality maps the 
twisted Cardy states in the former model to the states
in the latter which are associated with the conformal embedding  
$\s\uu (3)_2 \oplus \s\uu (2)_3 \subset \s \uu (6)_1$.

We have restricted ourselves to the Kazama-Suzuki models $G/H$
where $\text{rank}\,\g= \text{rank}\,\h$ and $H$ contains a single 
$U(1)$ factor.
It is straightforward to extend our analysis to the other cases,
namely, models with $\text{rank}~\g \neq \text{rank}~\h$ or
those with more than one $U(1)$ factors. 
These models may admit extra automorphisms other than 
those considered in this paper.
It is interesting to examine the corresponding twisted boundary condition,
in particular its relation to N=2 SCA.

Another interesting problem is the issue of the geometrical 
interpretation of the twisted Cardy states.
For the N=2 minimal models, combined with the N=2 Liouville theory,
it has been shown~\cite{Lerche,LLS} that the regular Cardy states 
can be interpreted as the cycles in the ALE spaces.
It is important to clarify whether the Cardy states 
in the Kazama-Suzuki models have the similar interpretation 
for some noncompact varieties~\cite{Song,LW}.

We have seen in section \ref{sec:exceptional} that, 
in $SU(3)/SU(2)\times U(1)$ at level two, one can obtain
the Cardy states other than the regular ones although 
this model does not admit non-trivial induced automorphisms.
This fact implies that the model $SU(3)/SU(2)\times U(1)$ has 
an extra automorphism which is not obtained from that of $SU(3)$.
In order to have all the automorphisms of the coset theory,
it is therefore insufficient to consider only the 
induced automorphisms.
It is important for the complete analysis of the boundary states
to find a systematic way of classifying all the automorphisms 
of the coset theory.

\section*{Acknowledgement}
We would like to thank H.~Awata, K.~Ito, S.~Mizoguchi,
T.~Nakatsu and H.~Suzuki for helpful discussions.
The work of H.~I. is supported in part by the Grant-in-Aid for Young 
Scientists from the Ministry of 
Education, Culture, Science, Sports and Technology of Japan. 
The work of T.~T. is supported by the Grant-in-Aid for Scientific
Research on Priority Areas (2) 14046201 from the Ministry of 
Education, Culture, Science, Sports and Technology of Japan.

\appendix
\section{Identification current groups for twisted Cardy states}
\label{sec:twistid}

In this appendix, we give the definition of the identification current 
groups $\Gid(\I^{\omega})$ and $\Gid(\tilde{\I})$, 
which are necessary for determining 
the set $\hat{\tilde{\I}}$ labeling the twisted Cardy states in the coset 
theory $G/H$ \eqref{eq:cosettwistedCardy}.
See \cite{IshikawaTani} for the detail. 

The simple current $J \in \G^G$ has a natural action 
on the labels of the Cardy states and the Ishibashi states.
In terms of the boundary state coefficients $\tilde{S}^G$,
this action can be written as follows:
\beq
\begin{split}
    &\tilde{S}^G_{\tilde{\alpha}J\lambda}=\tilde{b}^G_{\tilde{\alpha}}(J)
                                \tilde{S}^G_{\tilde{\alpha}\lambda}, \quad
                                                  J \in G(\I^{\omega^{\g}}),\\
    &\tilde{S}^G_{J\tilde{\alpha}\lambda}=\tilde{S}^G_{\tilde{\alpha}\lambda}
                                            b^G_{\lambda}(J),  \quad
                                                    J \in G(\tilde{\I}^G).
\label{eq:braneselectid}
\end{split}
\eeq 
Here $G(\I^{\omega^{\g}})$ and $G(\tilde{\I}^G)$ are the 
groups of the simple currents for the twisted Ishibashi states and 
the twisted Cardy states in the $G$ theory, respectively,
\beq
\begin{split}
   & G(\I^{\omega^{\g}})=\{J\in G_{\text{sc}}^G |\, 
                           J:\,\I^{\omega^{\g}}\mapsto \I^{\omega^{\g}}\}, \\
   & G(\tilde{\I}^G)=G_{\text{sc}}^G/{\cal S}(\tilde{\I}^G),
                \quad ~~
     {\cal S}(\tilde{\I}^G)=\{J_0 \in G_{\text{sc}}^G |\, 
                              J_0 \tilde{\alpha}=\tilde{\alpha}, 
                              \forall \tilde{\alpha} \in \tilde{\I}^G\}.
\label{eq:twistsimple}
\end{split}
\eeq 
One can also define these groups for the $H$ theory
in the same way as above, which we denote by $G(\I^{\omega^{\h}})$
and $G(\tilde{\I}^H)$.
The groups $\Gid(\I^{\omega})$ and 
$\Gid(\tilde{\I})$ of the identification currents for the 
twisted Ishibashi states and the twisted Cardy states 
in the coset theory are then defined as follows:
\beq
\begin{split}
  & \Gid(\I^{\omega})
                 =\Gid\cap (G(\I^{\omega^{\g}})\times G(\I^{\omega^{\h}})),\\
  & \Gid(\tilde{\I})=\Gid/(\Gid\cap\widehat{{\cal S}}(\tilde{\I})),        \\
  &   \widehat{{\cal S}}(\tilde{\I})=\{(J,J')\in \G\,|\, 
               b_{\lambda}^G(J)=b_{\mu}^H(J')~\forall (\lambda,\mu)\in 
               \I^{\omega^{\g}} \otimes \I^{{\omega}^{\h}}\}.  
\label{eq:cosetid}
\end{split}
\eeq

\section{N=2 superconformal algebra in terms of superfields}
\label{sec:KSsuper}

The N=2 superconformal algebra consists of the stress-energy 
tensor $T$, two supercurrents $G^{\pm}$ and the $U(1)$ current
$J$.
The N=1 supercurrent $G$ is given by a particular combination of 
$G^{\pm}$,
\beq
  G = \frac{1}{\sqrt{2}}(G^+ + G^-).
\eeq
The super stress-energy tensor $\T$ is then written as
\beq
    \T(z,\theta)  = \frac{1}{2} G(z) + \theta T(z)
                  = \frac{1}{2\sqrt{2}}(G^+ + G^-)(z)+\theta T(z). 
\eeq
The remaining generators of the N=2 SCA are arranged in a superfield $\GG$,
\beq
 \GG(z,\theta) = J(z) + \theta \frac{1}{\sqrt{2}}(-G^+ +G^-)(z).
\eeq 
In terms of these superfields, the N=2 SCA can be written as follows,
\bea
  & \T(z_1,\theta_1) \T(z_2,\theta_2) 
               \sim \frac{1}{z_{12}^3}\frac{c}{6}
                   +\frac{\theta_{12}}{z^2_{12}}\frac{3}{2}\T(z_2,\theta_2)
                   +\frac{1}{z_{12}}\frac{1}{2}D \T(z_2,\theta_2)
                   +\frac{\theta_{12}}{z_{12}}\partial \T(z_2,\theta_2),
                                     \nonumber \\
  & \T(z_1,\theta_1)\GG(z_2,\theta_2)
       \sim \frac{\theta_{12}}{z^2_{12}}\GG(z_2,\theta_2)
           +\frac{1}{z_{12}}\frac{1}{2}D \GG(z_2,\theta_2)
           +\frac{\theta_{12}}{z_{12}}\partial \GG(z_2,\theta_2),   \\
\label{eq:GandG}
  & \GG(z_1,\theta_1)\GG(z_1,\theta_1)
       \sim \frac{1}{z^2_{12}}\frac{c}{3}
           +\frac{\theta_{12}}{z_{12}}2\T(z_2,\theta_2). \nonumber      
\eea
The second equation shows that $\GG$ is a superprimary field 
with weight one.

\section{Uniqueness of the complex structure}
\label{sec:uniqueness}
We present the proof of the uniqueness of the complex structure 
$h_{\bar{a}\bar{b}}$ for the Kazama-Suzuki models
$G/H'\times U(1)$ with $\text{rank}\,\g = \text{rank}\,\h'+1$. 

Let $\bar{\Delta}$ be the set of the roots belonging to 
$\g\smallsetminus (\h'\oplus \uu(1))$.
This set is decomposed as
\beq
  \bar{\Delta}= \bar{\Delta}_+ \oplus \bar{\Delta}_-,
\eeq
where $+$ ($-$) means that the elements of $\bar{\Delta}_{\pm}$
are positive (negative) roots of $\g$.
We will show that, in the Cartan-Weyl basis, 
$h=(h^{\bar{a}}_{~\bar{b}})$ is a diagonal matrix with 
entries $+i$ for $\bar{\Delta}_+$ and 
$-i$ for $\bar{\Delta}_-$ (up to overall sign)
\footnote{In \cite{KS2}, this fact is used implicitly 
to derive the corollary from the theorem 1. 
Here we will give an explicit proof.}.
This means the uniqueness of the complex structure up to sign. 

The sets $\bar{\Delta}_{\pm}$ are further decomposed 
according to the values of the $U(1)$ charge.
As is shown in \cite{KS2}, the $U(1)$ charge is positive (negative)
for the roots in $\bar{\Delta}_{+}$ ($\bar{\Delta}_{-}$) and 
zero for the roots of $\h'$. 
The simple roots of $\g$ are consisted of the simple roots of $\h'$ and
$\alpha_{\times}$, which is the simple root of $\g$
with non-vanishing $U(1)$ charge.
Therefore, a root in $\bar{\Delta}_{\pm}$ has a $U(1)$ charge 
which is an integer multiple of the $U(1)$ charge 
for $\alpha_{\times}$. This yields the following decompositions, 
\beq
\begin{split}
  \bar{\Delta}_{+} &= \bar{\Delta}^{(1)}+ \bar{\Delta}^{(2)}+ \cdots , \\
  \bar{\Delta}_{-} &= \bar{\Delta}^{(-1)}+ \bar{\Delta}^{(-2)}+ \cdots , 
\end{split}
\eeq
where $\bar{\Delta}^{(n)}$ contains all the roots of $\g$ whose $U(1)$
charge is $n$ times that for $\alpha_{\times}$.
For the HSS models, $\bar{\Delta}^{(n)}$ is empty for 
$|n|\ge 2$.
For non-HSS models, however, this does not hold in general,
{\it e.g.}, $\bar{\Delta}^{(n)}$ is not 
empty for $-3\leq n \leq 3$ in $G_2/A_1^{>}\oplus \uu(1)$. 
An important property of the set $\bar{\Delta}^{(n)}$
is that it forms an irreducible representation of $\h'$.
This is because all the roots in $\bar{\Delta}^{(n)}$ are 
in the same conjugacy class of $\h'$ and the multiplicity of 
the adjoint representation of $\g$ is one.   
Based on this structure of $\bar{\Delta}$,
the proof proceeds in two steps:   
\begin{enumerate}
\item[(i)]
 $h$ is $\pm i$ on each $\bar{\Delta}^{(n)}$. 
\item[(ii)]
 $h=+i$ on $\bar{\Delta}^{(n)}$ and $-i$ on $\bar{\Delta}^{(-n)}$ for 
{\it any} positive $n$ (up to overall sign).  
\end{enumerate}
To show (i), we use the second condition in (\ref{eq:KSconditions0}).
In the matrix form $(f^a)^{\bar{b}}_{~\bar{c}}={f^{a\bar{b}}}_{\bar{c}}$, 
this condition is $f^a h = h f^a$ for all $a \in \h'$.
From Schur's lemma, we see that $h$ is proportional to the unit matrix
on each $\bar{\Delta}^{(n)}$.
The entries of $h$ are $+i$ or $-i$ by the first condition 
in (\ref{eq:KSconditions0}),
\beq
   h^{\bar{a}}_{~\bar{b}} = i\, h_n\, \delta^{\bar{a}}_{~\bar{b}} ,
                  \quad (\bar{a} \in \bar{\Delta}^{(n)}),
\eeq 
where $h_n=\pm 1$.
Then, one can prove (ii) if we show $h_{n}=-h_{-n}$ and $h_{n+1}=h_{n}$ for 
positive $n$.
For a diagonal $h$, the second condition in (\ref{eq:KSconditions0}) for $a=0$
reads,
\beq
   {f^{\bar{b}\bar{d}}}_{0} h^{\bar{d}}_{~\bar{d}}=
                      - h^{\bar{b}}_{~\bar{b}}{f^{\bar{b}\bar{d}}}_0.
\label{eq:pr1}
\eeq
Suppose $\bar{b}\in \bar{\Delta}^{(n)}$.
Since ${f^{\bar{b}\bar{d}}}_{0}$ is non-zero if and
only if ${\bar{b}}=-{\bar{d}}$, the equation (\ref{eq:pr1})
means that $h_{-n}=-h_{n}$.
For $h_{n+1}=h_n$, let $({\bar{a}},{\bar{b}},{\bar{c}}) \in 
(\bar{\Delta}^{(1)},\bar{\Delta}^{(n)},\bar{\Delta}^{(n+1)})$ and 
rewrite the third condition in (\ref{eq:KSconditions0}) as
\beq
   {f^{\bar{a}\bar{b}}}_{\bar{c}} = (h^{\bar{a}}_{~\bar{a}}h^{\bar{b}}_{~\bar{b}}
                                    -h^{\bar{b}}_{~\bar{b}}h^{\bar{c}}_{~\bar{c}}
                                    -h^{\bar{c}}_{~\bar{c}}h^{\bar{a}}_{~\bar{a}})
                                     {f^{\bar{a}\bar{b}}}_{\bar{c}}.
\eeq
We can choose a combination of $\bar{a}$, $\bar{b}$ and $\bar{c}$ such that
${f^{\bar{a}\bar{b}}}_{\bar{c}}\neq 0$ if the set $\bar{\Delta}^{(n+1)}$
is not empty. Then the above equation reads,
\beq
   1= -h_1 h_{n} +h_{n} h_{n+1} +h_{n+1}h_1.
\eeq 
Setting $n=1$, we find $h_1 h_2=1$, which means $h_2=h_1$.
By the induction, we find that $h_{n}= h_1$ for all positive $n$.
This completes the proof.

\section{Selection rules and field identifications}
\label{sec:selectid}
In this appendix, we briefly review
the selection rules and the field 
identifications~\cite{MS,Gepner,LVW,Schellekens}
for the models considered in the present paper.
\subsection{$SU(m+n)/SU(m)\times SU(n)\times U(1)$}
\label{sec:suselectid}

The bosonic form of this model is given in \eqref{eq:mnk}. 
A primary field is labeled as $(\Lambda,\tilde{\Lambda}; \lambda ,\sigma)$ 
where
\bea
 \Lambda & =(\Lambda_1,\Lambda_2,\ldots,\Lambda_{m+n-1})
            \in P_+^{k}({\mathfrak s}\uu(m+n)),                \nonumber  \\
 \tilde{\Lambda}
         & \in P_{+}^{1}(\s\oo (2mn)) = \{o,v,s,c\},           \nonumber \\
 \lambda & = (\lambda^{(1)},\lambda^{(2)})               
           = ((\lambda^{(1)}_1,\lambda^{(1)}_2,\ldots,\lambda^{(1)}_{m-1}),
              (\lambda^{(2)}_1,\lambda^{(2)}_2,\ldots,\lambda^{(2)}_{n-1})) 
                                                                \\
         &  \hspace{6cm} \in P_+^{n+k}({\mathfrak s}\uu(m))\times 
                P_+^{m+k}({\mathfrak s}\uu(n)),                \nonumber  \\
 \sigma &  \in P_+^{K}(\uu (1))=\{ 0,1,\dots, K-1\}  
                             \quad (K\equiv mn(m+n)(m+n+k)). 
                                                               \nonumber
\eea
The selection rule is given by 
\beq
\begin{split}
  m\,r_{\Lambda} +  mn(m+n)\,\epsilon_{\tilde{\Lambda}}/2 
         =   (m+n)\, r_{\lambda^{(1)}} + \sigma &\quad~~ \text{mod}~~m(m+n),  \\
  n\, r_{\Lambda} +  mn(m+n)\,\epsilon_{\tilde{\Lambda}}/2
         =   (m+n)\, r_{\lambda^{(2)}} - \sigma &\quad~~ \text{mod}~~n(m+n),
\end{split}
\label{eq:selection}
\eeq
where
\beq
\begin{split}
  &  r_{\Lambda}= \sum_j j\Lambda_j,~~
     r_{\lambda^{(i)}}=\sum_j j\lambda^{(i)}_j,          \\
  &  \epsilon_{\tilde{\Lambda}} = \left\{
                \begin{array}{ll}
                0 & \quad \tilde{\Lambda}=o,v~~(\text{NS}),       \\
                1 & \quad \tilde{\Lambda}=s,c~~(\text{R}).
                \end{array} 
                \right.
\end{split}
\eeq
Correspondingly, the identification group $\Gid$
has the following two generators,
\beq
\begin{split}
  J_{(1)} &= (J,\tilde{J}_v^n\,/\,(J',1),-n(m+n+k)),           \\
  J_{(2)} &= (J,\tilde{J}_v^m\,/\,(1,J'),m(m+n+k)),
\label{eq:suidcurrent}
\end{split}
\eeq
where each simple currents act as follows, 
\beq
\begin{split} 
  & J:~~\Lambda_j \rightarrow \Lambda_{j-1},\quad \quad
    \tilde{J}_v:~~(o,v,s,c)\mapsto (v,o,c,s),                 \\
  & J':~~\lambda^{(i)}_j \rightarrow \lambda^{(i)}_{j-1}, \quad ~~
    p:~~\sigma \mapsto \sigma + p~~~(p \in \{0,1,\dots,K-1\}).
\end{split}
\eeq
The monodromy charges are given by
\beq
\begin{split}
 &  b_{\Lambda}(J)=e^{2\pi i~ r_{\Lambda}/(m+n)},\quad   
   b_{\tilde{\Lambda}}(\tilde{J}_v)=e^{2\pi i~ \epsilon_{\tilde{\Lambda}}/2},\\
 &  b_{\lambda^{(1)}}(J') =e^{2\pi i~ r_{\lambda^{(1)}}/m},~~
    b_{\lambda^{(2)}}(J') =e^{2\pi i~ r_{\lambda^{(2)}}/n}, \quad
    b_{\sigma}(p)=e^{-2\pi i~ p\, \sigma/(mn(m+n)(m+n+k))}.
\end{split}
\eeq
The identification current group has no fixed points unless $m$, $n$ and $k$
have a common divisor.
%

\subsection{$SO(2n)/SO(2n-2)\times U(1)$}
\label{sec:soselectid}

The bosonic form of this model is given in \eqref{eq:SO}.
The selection rule is given by
\beq
\begin{split}
  & r^{(v)}_{\Lambda} = r^{(v)}_{\lambda}  \hspace{4.1cm}(\text{mod}~~1),    \\
  & r^{(s)}_{\Lambda}+(n-1) \epsilon_{\tilde{\Lambda}} /2
   =r^{(s)}_{\lambda}+\sigma/4             \quad  (\text{mod}~~1).         
\end{split}
\eeq
Here $r^{(s)}_{\Lambda}$ and $r^{(v)}_{\Lambda}$ are the inner 
products of the weight $\Lambda$ with the spinor weight $\Lambda_s$ 
and the vector weight $\Lambda_{v}$, respectively,
\beq
\begin{split}
  r^{(s)}_{\Lambda} &= (\Lambda,\Lambda_s)
                     =\frac{1}{2}\{ \Lambda_1+ 2 \Lambda_2+ \cdots 
                                +(n-2)\Lambda_{n-2}+\frac{n-2}{2}\Lambda_{n-1}
                                +\frac{n}{2}\Lambda_{n}\} ,          \\
  r^{(v)}_{\Lambda} &= (\Lambda,\Lambda_v)
                     = \Lambda_1+\Lambda_2+\cdots +\Lambda_{n-2}
                                +\frac{1}{2} \Lambda_{n-1}
                                +\frac{1}{2} \Lambda_{n}.
\label{eq:selectionso}
\end{split}
\eeq
$r^{(s)}_{\lambda}$ and $r^{(v)}_{\lambda}$ are defined similarly.
The field identification group is generated by 
\beq
\begin{split}
  J_{(1)} &= (J_{v},1\,/\,J'_v,0),                            \\
  J_{(2)} &= (J_{s},\tilde{J}_{v}^{n-1}\,/\,J'_{s},k+2(n-1)),
\label{eq:fieldidso}
\end{split}
\eeq  
where $J_{\Lambda}$ is the simple current of
the representation $\Lambda$.

\section{Explicit form of the twisted Cardy states}
\label{sec:explicit}
In this appendix, we give the explicit form of the twisted 
Cardy states for the two examples considered in section \ref{sec:example}.
\subsection{$SU(4)/SU(2)\times SU(2)\times U(1)$}
\label{sec:suexplicit}

\begin{figure}[tb]
\begin{center}
\includegraphics[width=1.8cm]{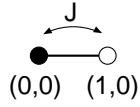}
\end{center}
\caption{The twisted NIM-rep graph $n_{(1,0,0)}$ of the $SU(4)_1$
WZW model.
The black (white) node represents $\tilde{b}(J^2)=1 (-1)$.}
\label{fig:su4}
\end{figure}

\begin{figure}[tb]
\begin{center}
\includegraphics[width=3.2cm]{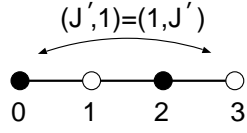}
\end{center}
\caption{The twisted NIM-rep graph $n_{(1,0)}=n_{(0,1)}$
of the $SU(2)_3\times SU(2)_3$ WZW model.
The black (white) nodes represent $\tilde{b}(J',J')=1 (-1)$ 
($(J',J') \in G({\I}^{\pi})$). The action of the simple current
$(J',1)=(1,J') \in G({\tilde{\I}})$ is also shown.} 
\label{fig:su2su2}
\end{figure}

In this case, the non-trivial automorphism is written as 
$(\omega_o,1;\pi,1)$ (see \eqref{eq:su4su2su2lift}).
We first construct twisted Cardy states in  
two sectors $SU(4)_1$ and $SU(2)_3\times SU(2)_3$.
From the formula given in section \ref{sec:WZWmutual},
we obtain the following form of the boundary state 
coefficients $\tilde{S}$ and the simple current groups $G(\I^{\omega})$
and $G(\tilde{\I})$:
\bigskip
\\
\underline{$SU(4)_1$}          
\medskip
\\
The boundary state coefficients are given by the modular 
transformation matrix for the twisted chiral algebra $A_3^{(2)}$
at level one,
\beq
\begin{split}
  & \tilde{S}^{SU(4)_1}  = \frac{1}{\sqrt{2}}
                  \left(
                  \begin{array}{cc}
                  1 & 1   \\
                  1 &-1
                  \end{array} 
                  \right),                               \\
  &\tilde{\I} = P_+^1(A_3^{(2)})=\{(0,0),(1,0)\},
~~\I^{\omega_o} = P_+^1(D_3^{(2)})=\{(0,0),(0,1)\}.
\end{split}
\eeq
The simple current group of this theory is
\beq
    \G = \{ J^j |\, j =  0,1,2,3 \}\cong \Z_4.
\eeq 
The simple current groups for $\I^{\omega_o}$ and $\tilde{\I}$ are 
\beq
\begin{split}
   & G(\I^{\omega_o}) = \{1,J^2\}\cong \Z_2,           \\
   & G(\tilde{\I}) = \{1,J\}\cong \Z_2, 
                    \quad ~~ {S}(\tilde{\I})=\{1,J^2\}.
\end{split}
\eeq
We show the action of the simple current $J$ of $G(\tilde{\I})$
and $\tilde{b}(J^2)$ ($J^2 \in G(\I^{\omega_o})$)
in Fig.~\ref{fig:su4}.
%
%
\bigskip
\\
\underline{$SU(2)_3 \times SU(2)_3$}
\medskip
\\
The boundary state coefficients are given by the modular 
transformation matrix for $SU(2)_3$,
\beq
\begin{split}
   &\tilde{S}^{SU(2)_3\times SU(2)_3} = S^{SU(2)_3} ,       \\
   &\tilde{\I}=\I=\{\tilde{\lambda}\,|\, \tilde{\lambda} = 0,1,2,3\},~~
    \I^{\pi} = \{(\lambda,\lambda)\,|\,
                                         \lambda=0,1,2,3\}, 
\end{split}
\eeq
and the simple current groups read
\beq
\begin{split}
   & \G= \{({J'}^i,{J'}^j)|\,i,j =0,1 \} \cong \Z_2 \times \Z_2,\\
   & G(\I^{\pi}) = \{ ({J'}^k,{J'}^k)\in \G
                      |\, k =0,1 \} \cong \Z_2,          \\
   & G(\tilde{\I}) = \G/{\cal S}(\tilde{\I}) \cong \Z_2 ,
                    \quad ~~  {\cal S}(\tilde{\I})
                             =\{({J'}^k,{J'}^k) |\, k=0,1\}.
\end{split}
\eeq
We show the action of these groups in Fig.~\ref{fig:su2su2}.
%
%
\bigskip

Then, we can construct twisted Cardy states 
in the coset theory by applying the procedure reviewed in 
section \ref{sec:cosetmutual} and in appendix \ref{sec:twistid}.
The set $\hat{\I}^{\hat{\omega}_o}$ (\ref{eq:cosettwistspec}) is given by
\beq
  \hat{\I}^{\hat{\omega}_o}=\{ ((0,0),\tilde{\Lambda};\,(0,0),8j),
                  ((0,0),\tilde{\Lambda};\,(1,1),4+8j)|\,
                      \tilde{\Lambda}\in \{o,v,s,c\};~ j=0,1,\dots,4\}.
\eeq
From the definitions (\ref{eq:cosetid}), the identification current 
groups read 
\beq
\begin{split}
 & \Gid(\I^{\omega})= \{(J_{(1)}J_{(2)}^{-1})^{k}(J_{(1)}J_{(2)})^{k'}|\, 
                      k=0,1,2,3;\,k'=0,1\} \cong\Z_4\times \Z_2,  \\
 & \Gid(\tilde{\I})=\{J_{(1)}^i=J_{(2)}^{-i}|\, i=0,1,\dots,7\}
                         \cong \Z_8, \quad ~~ 
   \Gid \cap \widehat{{\cal S}}(\tilde{\I})
                                   = \{(J_{(1)}J_{(2)})^{k'} |\, k'=0,1\}.
\end{split}
\eeq
The labels (\ref{eq:cosettwistedCardy}) 
of 40 twisted Cardy states are given as follows:
\beq
  \hat{\tilde{\I}}=\{((0,0),\tilde{\Lambda};\, 0,4j),
                 ((0,0),\tilde{\Lambda};\, 2,4j)|\,
                     \tilde{\Lambda}\in \{o,v,s,c\};~j=0,1,\dots,4\}.
\eeq
The boundary state coefficients $\hat{\tilde{S}}$ (\ref{eq:cosettwisted})
are given by
\beq
  \hat{\tilde{S}} = 8 \, \tilde{S}^{SU(4)_1} S^{SO(8)_1}
                     \overline{ \tilde{S}^{SU(2)_3\times SU(2)_3} 
                                S^{U(1)_{80}}},
\label{eq:KSperm}
\eeq
since $|\Gid(\I^{\omega})|=|\Gid(\tilde{\I})|=8$.

\subsection{$SO(8)/SO(6)\times U(1)$}
\label{sec:soexplicit}

\begin{figure}[tb]
\begin{center}
\includegraphics[width=2.3cm]{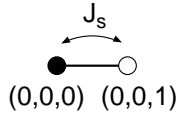}
\end{center}
\caption{The twisted NIM-rep graph $n_{(0,0,0,1)}$ of the $SO(8)_1$
WZW model.
The black (white) node represents $\tilde{b}(J_v)=1 (-1)$.}
\label{fig:so8}
\end{figure}

\begin{figure}[tb]
\begin{center}
\includegraphics[width=3.7cm]{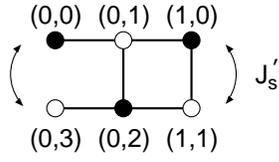}
\end{center}
\caption{The twisted NIM-rep graph $n_{(0,0,1)}$ in the
$SO(6)_3$ theory. 
The black (white) nodes represent
$\tilde{b}({J'}_s^{2})=1$ ($-1$). The action of the 
simple current $J'_s$ is also shown.}
\label{fig:twistAD}
\end{figure}

In this case, the non-trivial automorphism is written as
$(\omega_2,1;\omega_c',1)$ (see (\ref{eq:solift})).
The twisted Cardy states in the $SO(8)_1$ and the 
$SO(6)_3$ sectors are as follows:
\bigskip
\\
\underline{$SO(8)_1$}
\medskip
\\
The boundary state coefficients $\tilde{S}$ is the modular 
transformation matrix of $D_4^{(2)}$ at level one,
\bea
  & \tilde{S}^{SO(8)_1} = \frac{1}{\sqrt{2}}
                  \left(
                  \begin{array}{cc}
                  1 & 1   \\
                  1 &-1
                  \end{array} 
                  \right),                               \\
   &\tilde{\I} = P_+^1(D_4^{(2)})=\{(0,0,0),(0,0,1)\}, 
~~ \I^{\,\omega_2} = P_+^1(A_5^{(2)})=\{(0,0,0),(1,0,0)\}.  \nonumber 
\eea
The simple current groups for $\I^{\omega_2}$ and $\tilde{\I}$ read
\beq
\begin{split}
   & \G =\{ J_s^i J_v^ j |\, i,j =0,1\}\cong \Z_2\times \Z_2, \\
   & G(\I^{\omega_2}) = \{1,J_v \} \cong \Z_2,  \\
   & G(\tilde{\I}) = \{1,J_s \} \cong \Z_2.
\end{split}
\eeq
The action of these groups are shown in Fig.~\ref{fig:so8}.
%
\bigskip
\\
\underline{$SO(6)_3$}
\medskip
\\
The boundary state coefficients $\tilde{S}$ is the modular 
transformation matrix of $D_3^{(2)}$ at level three, 
\bea
   & \tilde{S}^{SO(6)_3}_{\tilde{\lambda}\lambda}
               = \left(
                     \begin{array}{rrr}
                     x \,\kappa  & -y \,\kappa & -z \,\kappa  \\
                     -y \,\kappa & z \,\kappa  & x \,\kappa   \\
                     -z \,\kappa  & x \,\kappa  & y \,\kappa      
                     \end{array}
                     \right),                                            \\
   &  \quad \quad ~\kappa \equiv \frac{1}{\sqrt{2}}\left(
                                            \begin{array}{rr}
                                            1 & 1 \\
                                            1 & -1 
                                            \end{array}
                                            \right),           \nonumber    \\
   &  \quad  \quad ~ x = \frac{2}{7}(2 c_2 - c_6 -1),~
           y = \frac{2}{7}(2 c_6 -c_4 -1),~
           z = \frac{2}{7}(2 c_4 -c_2 -1)
     \quad (c_n \equiv \text{cos}\frac{n \pi}{7}),         \nonumber \\
   & \tilde{\I} = P_+^{3}(D_{3}^{(2)}) 
              =\{\tilde{\lambda}=(\tilde{\lambda}_1,\tilde{\lambda}_2)\}
              =\{(0,0),(0,3),(0,2),(0,1),(1,0),(1,1)\},          \nonumber \\
   & {\I}^{\,\omega'_c} = P_+^{3}({A}_3^{(2)})
              =\{\lambda=(\lambda_1,\lambda_2=\lambda_3)\}
              =\{(0,0),(3,0),(2,0),(1,0),(0,1),(1,1)\}.    \nonumber
\eea
The simple current groups are given by
\beq
\begin{split}
  & \G= \{{J'_s}^k |\, k =0,1,2,3\} \cong \Z_4,                     \\
  & G({\I}^{\,\omega'_c}) = \{1,{J'}_s^2=J'_v \} \cong \Z_2,           \\
  & G(\tilde{\I}) = \{1,J'_s\} \cong \Z_2, \quad ~~
                        {\cal S}(\tilde{\I}) = \{1,{J'}_s^2\}.
\end{split}
\eeq
The action of these groups is shown in Fig.\ref{fig:twistAD}.
%
\bigskip

The simple current groups for the coset theory are given by
\beq
\begin{split}
  &\Gid({\I}^{\omega}) = \{J_{(1)}^i,\,J_{(2)}^{2j}\,|\, i,j=0,1\}
                          \cong \Z_2 \times \Z_2 ,                     \\
  &\Gid(\tilde{\I})  = \{J_{(2)}^k |\, k=0,1,2,3\}\cong \Z_4,
                \quad \quad   
   \Gid \cap \widehat{\cal S}(\tilde{\I}) = \{1,J_{(1)}\}.
\end{split}
\eeq 
The boundary state coefficients $\hat{\tilde{S}}$ (\ref{eq:cosettwisted}) 
are then written as 
\beq
\begin{split}
  & \hat{\tilde{S}} = 4 \, \tilde{S}^{SO(8)_1}S^{SO(12)_1} 
                \overline{\tilde{S}^{SO(6)_3} S^{U(1)_{28}}},         \\   
  & \hat{\I}^{\hat{\omega}_2}= \{(o,\tilde{\Lambda}_{\text{NS}}\,;\,
                              (\lambda_1,\lambda_2),4j),
                   (o,\tilde{\Lambda}_{\text{R}}\,;\,
                              (\lambda_1,\lambda_2),4j+2)      \\
                 &\hspace{1cm}|~ \tilde{\Lambda}_{\text{NS}}\in \{o,v\},
                                 \tilde{\Lambda}_{\text{R}}\in \{s,c\};\,
                 (\lambda_1,\lambda_2) 
                \in \{(0,0),(2,0),(0,1)\};~j=0,1,\dots, 6\},                \\
  & \hat{\tilde{\I}} =\{(o,\tilde{\Lambda}\,;\, 
                   (\tilde{\lambda}_1,\tilde{\lambda}_2),2j)\,\\
                 &\hspace{3cm} |~    \tilde{\Lambda}\in \{o,v,s,c\};\,
                       (\tilde{\lambda}_1,\tilde{\lambda}_2) 
                        \in \{(0,0),(0,2),(1,0)\};~
                        j=0,1,\dots,6\}. 
\end{split}
\eeq
In this way, we obtain 84 A-type twisted Cardy states.


\end{document}